\documentclass[aps,prc,12pt,nofootinbib]{revtex4}

\usepackage{amsfonts}
\usepackage{amsmath,epsfig,bm}
\usepackage{graphicx}
\usepackage{color}

\usepackage{tabularx} 

\usepackage{times}

\newcommand{\be}{\begin{equation}}
\newcommand{\ee}{\end{equation}}
\newcommand{\ba}{\begin{eqnarray}}
\newcommand{\ea}{\end{eqnarray}}
\newcommand{\bd}{\begin{displaymath}}
\newcommand{\ed}{\end{displaymath}}

\def\rt3{\sqrt{3}}
\def\rt6{\sqrt{6}}
\def\mL2{(m_{\phi}L)^2}

\begin{document}

\title{{\bf Large Baryon Densities Achievable in High Energy Heavy Ion Collisions Outside the Central Rapidity Region}}

\author{Ming Li and Joseph I. Kapusta}
\affiliation{School of Physics and Astronomy, University of Minnesota, Minneapolis Minnesota 55455, USA}

\vspace{.3cm}
\date{\today}

\parindent=20pt

\begin{abstract}

Nuclei are nearly transparent to each other when they collide at high energy, but the collisions do produce high energy density matter in the central rapidity region where most experimental measurements are made. What happens to the receding nuclear fireballs? We calculate the energy loss of the nuclei using the color glass condensate model. We then use a simple space-time picture of the collision to calculate the baryon and energy densities of the receding fireballs. For central collisions of large nuclei at the BNL Relativistic Heavy Ion Collider  and the CERN Large Hadron Collider we find baryon densities more than ten times that of normal nuclear matter. These results provide initial conditions for subsequent hydrodynamic evolution and could test the equation of state at very high baryon densities.

\end{abstract}

\maketitle

\section{Introduction}

Since the 1980s the high-energy heavy-ion community has focused on the central rapidity region of nucleus-nucleus collisions \cite{QMseries}.  That followed the influential work of Bjorken in 1983 \cite{BJ1983}.  The reasons are: (i) the energy density is expected to be higher there; (ii) the matter is nearly baryon-free, making it relevant to the type of matter that existed in the early universe and more amenable to comparisons with lattice gauge calculations; (iii) detectors in a collider can more readily measure particle production and correlations within one unit of rapidity around the center-of-momentum of the colliding nuclei.  The earlier work of Anishetty, Koehler and McLerran in 1980 \cite{AKM1980}, which found that nuclei were significantly compressed and excited when they collide at extreme relativistic energies, was pursued only sporadically.  In Ref. \cite{LKRapid} we picked up the line of work of Ref.
\cite{AKM1980} in light of the theoretical and experimental developments since then.  We found that in central collisions of gold nuclei at $\sqrt{s_{NN}} = 200\,\rm{GeV}$, the top BNL Relativistic Heavy Ion Collider (RHIC) energy, the central core lost about $\Delta y = 2.9$ units of rapidity and was compressed by a factor of ${\rm e}^{\Delta y}$ \cite{AKM1980,C1984,GC1986,LKRapid}, a very large compression indeed.  In this paper we expand our work on the baryonic fireballs which emerge beyond one unit of rapidity.  It should be noted immediately that our calculations are not particularly relevant to the lower beam energy scan at RHIC.

In our work we use the semi-analytical results of Refs. \cite{CFKL2015,LK2016}, which were based on the McLerran-Venugopalan model \cite{McLerranVenu}, for the energy and momentum deposited in the central rapidity region by the colliding nuclei.  Similar numerical results were obtained in Refs. \cite{Lappi:2006fp,Fukushima:2011nq,Gelis:2013rba}.
This produced matter is oftentimes referred to as glasma; it is the precursor to quark-gluon plasma. We consider the backreaction of the glasma on the receding nuclei. Imposing energy and momentum conservation on the whole system, the increase of energy and momentum for the glasma is equal to the decrease of energy and momentum of the nuclei. Solving the energy momentum conservation equations gives us the anticipated rapidity loss and excitation energy for the receding nuclear fireballs. We neglect transverse coupling over the brief time duration of the collision, which is less than 1 fm/c.  This means that the collisions are treated as a sum of independent tube-tube or streak-streak collisions, dependent on the transverse position $\mathbf{r}_{\perp}$.  Once we have the baryon density and energy density, we can estimate the resulting temperature and baryon chemical potential of the fireballs as functions of $\mathbf{r}_{\perp}$. This is done with the help of a realistic crossover equation of state that smoothly connects the quark-gluon plasma phase at high temperature and high baryon chemical potential to the hadronic resonance gas phase at low temperature and small baryon chemical potential \cite{Albright}.  Apart from the prototype Au+Au collision at $\sqrt{s_{NN}} = 200\,\rm{GeV}$, we also explore other collision configurations, such as the asymmetric Cu+Au collision and collisions involving distorted nuclei like U+U collisions. 
Non-central collisions are also studied. In addition, we consider Pb+Pb collisions at even higher energy at the CERN Large Hadron Collider (LHC).   
  
The outline of the paper is as follows. In Sec. \ref{section:high_baryon_density_formalism} the formalism for implementing energy-momentum conservation is presented. Properties of the equations of motion and determination of the input parameters are discussed. Section \ref{section: Au_Au200GeV} is devoted to the prototype Au+Au central collision at $\sqrt{s_{NN}} = 200\,\rm{GeV}$. In Sec. \ref{section:systematic_studies}, systematic studies of the high baryon densities in high energy heavy-ion collisions are given. We consider the nuclear size dependence, the collision energy dependence, and the impact parameter dependence.  Conclusions are presented in Sec. \ref{section:Conclusion}.

\section{Baryon Rapidity Loss and Nuclear Excitation Energy}
\label{section:high_baryon_density_formalism}

In this section we discuss the equations of motion and input parameters, followed by some numerical results.

\subsection{Equations of Motion}

For high energy heavy-ion collisions in the center-of-momentum frame, the nuclei are highly Lorentz contracted spheres (or ellipsoids if the ground state nucleus is deformed). We neglect transverse dynamics, which should not be important during the fraction of a fm/c time interval of relevance.  Then the collision can be thought of as a sum of tube-tube or streak-streak collisions, each taking place at a particular value of the transverse coordinate ${\bf r}_\perp$ with the beam along the $z$-axis.  The projectile tube has a four-momentum per unit area in the center-of-momentum frame denoted by ${\cal P}_{\rm P}^{\mu} = ({\cal E}_{\rm P}, 0, 0, {\cal P}_{\rm P})$.  The tube loses energy and momentum to the classical color electric and magnetic fields produced in the region between the two receding tubes.  This loss is quantified by 
\begin{equation}\label{eq:glasma_slab_EM_conservation}
d{\cal P}^{\mu}_{\rm P} = -T_{\rm glasma}^{\mu\nu}d\Sigma_{\nu}
\end{equation}
where $d\Sigma_{\nu} = (dz,0,0,-dt)$ is the infinitesimal four-vector perpendicular to the hypersurface spanned by $dt$, $dz$, and unit transverse area.  The energy-momentum tensor of the glasma has been calculated in Refs. \cite{CFKL2015,LK2016} as
\begin{equation}\label{eq:glasma_em_tensor}
T^{\mu\nu}_{\rm glasma}=
\begin{pmatrix}
\mathcal{A}+\mathcal{B}\cosh{2\eta} & 0 & 0 & \mathcal{B}\sinh{2\eta} \\
0 & \mathcal{A} & 0 & 0 \\
0 & 0 & \mathcal{A} & 0 \\
\mathcal{B}\sinh{2\eta} & 0 & 0 & -\mathcal{A}+\mathcal{B}\cosh{2\eta} \\
\end{pmatrix} \, .
\end{equation}
The $\mathcal{A}$ and $\mathcal{B}$ are functions of proper time $\tau =\sqrt{t^2-z^2}$ (and other input parameters to be discussed in the following), while the dependence on space-time rapidity $\eta=\frac{1}{2}\ln [(t+z)/(t-z)]$ follows from the fact that $T^{\mu\nu}_{\rm glasma}$ is a second-rank tensor in a boost-invariant setting.  The longitudinal position of the tube $z_{\rm P}$ is a function of time, $z_{\rm P}=z_{\rm P}(t)$.  The $z_{\rm P}$ is related to the time $t$ via the velocity $v_{\rm P} = dz_{\rm P}/dt =\tanh{y_{\rm P}}$, where $y_{\rm P}$ is the momentum-space rapidity of the tube. Hence all the quantities solely depend on $t$.  Note that $T^{\mu\nu}_{\rm glasma}$ must be evaluated on the trajectory of the tube.  Explicitly
\begin{equation}\label{eq:slab_eom}
\begin{split}
&d{\cal E}_{\rm P}(t,z_{\rm P}) = -T^{00}_{\rm glasma}(t,z_{\rm P})dz_{\rm P} + T^{03}_{\rm glasma}(t,z_{\rm P})dt\, , \\
&d{\cal P}_{\rm P}(t,z_{\rm P}) = -T^{30}_{\rm glasma}(t,z_{\rm P})dz_{\rm P} + T^{33}_{\rm glasma}(t,z_{\rm P})dt\, . \\
\end{split}
\end{equation}
It is useful to define the Lorentz invariant effective mass per unit area ${\cal M}_{\rm P}$ via the relations ${\cal E}_{\rm P} = {\cal M}_{\rm P} \cosh{y_{\rm P}}$ and ${\cal P}_{\rm P} = {\cal M}_{\rm P} \sinh{y_{\rm P}}$ so that
\begin{equation}\label{eq:intermediate_step_1}
\begin{split}
 &d\mathcal{E}_{\rm P} =  \cosh{y_{\rm P}}\, d\mathcal{M}_{\rm P} + {\cal M}_{\rm P} \sinh{y_{\rm P}}\, dy_{\rm P}\, ,\\
&d\mathcal{P}_{\rm P} = \sinh{y_{\rm P}} \,d\mathcal{M}_{\rm P} + {\cal M}_{\rm P} \cosh{y_{\rm P}}\, dy_{\rm P}\,. \\
\end{split}
\end{equation}
We then express the differential form of the energy momentum conservation in Eqs. \eqref{eq:slab_eom} in terms of the Milne coordinates $(\tau,\mathbf{x}_{\perp}, \eta)$. Using the transformations $\tau = \sqrt{t^2-z_{\rm P}^2}$ and $\eta_{\rm P} = \frac{1}{2} \ln [(t+z_{\rm P})/(t-z_{\rm P})]$, the pseudorapidity of the projectile slab $\eta_{\rm P}$ follows the equation
\begin{equation}\label{eq:eta_eom}
\tau \frac{d\eta_{\rm P}}{d\tau} = \tanh{(y_{\rm P} - \eta_{\rm P})}
\end{equation}
where the auxiliary relations between $(t, z_{\rm P})$ and $(\tau, \eta_{\rm P})$
\begin{equation}\label{eq:intermediate_step_2}
\begin{split}
&\frac{d\tau}{dt} = \frac{\cosh{(y_{\rm P} - \eta_{\rm P})}}{\cosh{y_{\rm P}}}\, ,\\
&\frac{d\tau}{dz_{\rm P}} = \frac{\cosh{(y_{\rm P} - \eta_{\rm P})}}{\sinh{y_{\rm P}}}\,,\\
\end{split}
\end{equation}
have been used. Substituting Eqs. \eqref{eq:glasma_em_tensor}, \eqref{eq:intermediate_step_1} and \eqref{eq:intermediate_step_2} into Eq. \eqref{eq:slab_eom}, we obtain the equations of motion for the rapidity $y_{\rm P}$ and the effective mass $\mathcal{M}_{\rm P}$ of the projectile slab in terms of Milne coordinates
\begin{equation}\label{eq:slab_yM_eom}
\begin{split}
&-\mathcal{M}_{\rm P} \cosh{(y_{\rm P} - \eta_{\rm P})} \frac{dy_{\rm P}}{d\tau} = \mathcal{A}(\tau) -\mathcal{B}(\tau) \cosh{(2y_{\rm P} - 2\eta_{\rm P})}\, ,\\
&\cosh{(y_{\rm P} - \eta_{\rm P})} \frac{d\mathcal{M}_{\rm P}}{d\tau} = -\mathcal{B}(\tau) \sinh{(2y_{\rm P} -2\eta_{\rm P})}\, . \\
\end{split}
\end{equation}
Note that Eq. \eqref{eq:slab_yM_eom} should be supplemented by the equation for pseudorapidity $\eta_{\rm P}$ via Eq. \eqref{eq:eta_eom}. It is important to point out that $y_{\rm P}$ and $\mathcal{M}_{\rm P}$ are dynamical variables while $\eta_{\rm P}$ is a geometric variable coming from the coordinate transformation. Equations  \eqref{eq:eta_eom} and \eqref{eq:slab_yM_eom} are coupled first order differential equations. Given the functions $\mathcal{A}(\tau)$ and $\mathcal{B}(\tau)$ from the glasma energy-momentum tensor, initial conditions are needed to solve for $y_{\rm P}$, $\mathcal{M}_{\rm P}$ and $\eta_{\rm P}$. But before  discussing the initial conditions and numerical solutions to the equations, a few important properties of Eqs. \eqref{eq:eta_eom} and \eqref{eq:slab_yM_eom} are worth noting. 
\begin{enumerate}
\item[(i)] If $\mathcal{B}(\tau) =0$, the glasma energy-momentum tensor becomes diagonal, $T^{\mu\nu}_{\rm glasma} = \rm{diag}(\mathcal{A},\mathcal{A},\mathcal{A},-\mathcal{A})$, which is like the string rope model \cite{MK2002,Shen:2017bsr}. In that case, $d\mathcal{M}_{\rm P}/d\tau =0$ and the effective mass $\mathcal{M}_{\rm P}$ does not change with time. No energy is deposited in the nucleus. On the other hand, the rapidity $y_{\rm P}$ decreases with time. As long as $\mathcal{A}(\tau) \neq 0$, $y_{\rm P}$ decreases until it becomes negative when the yo-yo type motion in string rope models appears. Therefore, a nonvanishing $\mathcal{B}(\tau)$ is necessary to incorporate nuclear excitation energy and prevent the appearance of yo-yo type motion. The glasma energy-momentum tensor predicts a nonvanishing $\mathcal{B}(\tau)$. Physically, the function $\mathcal{A}(\tau)$ represents contributions from the longitudinal chromo-electromagnetic fields $E^z$ and $B^z$, while the function $\mathcal{B}(\tau)$ represents contributions from the transverse chromo-electromagnetic fields $E^i$ and $B^i$. Initially, at $\tau=0$, $\mathcal{A} \neq 0$ and $\mathcal{B} =0$, so the transverse fields are zero while the longitudinal fields are nonzero. As the glasma evolves, $\mathcal{B}$ gradually increases while $\mathcal{A}$ gradually decreases until they become nearly equal; see Sect. IIc. 

\item[(ii)] If $\mathcal{B}(\tau) > 0$, the effective mass $\mathcal{M}_{\rm P}(\tau)$ increases or decreases with time according to the sign of $y_{\rm P}(\tau) - \eta_{\rm P}(\tau)$. For the excitation energy $\mathcal{M}_{\rm P}(\tau)$ to increase with time, $y_{\rm P}(\tau) < \eta_{\rm P}(\tau) $ has to be maintained. From the definition $\tanh{y_{\rm P}} = dz_{\rm P}/dt$, one obtains $z_{\rm P} = \int^{t}_0\, \tanh{y_{\rm P}(t^{\prime})} dt^{\prime}> \tanh{y_{\rm P}(t)}\, t$ because the velocity of the projectile tube $v_{\rm P} = \tanh{y_{\rm P}}$ decreases with time. With $\tanh{\eta_{\rm P}} = z_{\rm P}/t$, the condition $y_{\rm P}< \eta_{\rm P}$ is strictly guaranteed. Hence, positive values of $\mathcal{B}(\tau)$ predicts an increase of the effective mass and nuclear excitation energy. In addition, as long as $y_{\rm P} \neq \eta_{\rm P}$, the effective mass always increases with time, which brings up the question of a cutoff time dependence of the nuclear excitation energy.   Physically, the increase of effective mass is governed by the strength of the transverse chromo-electromagnetic fields in the glasma which gradually builds up over time. It is worthwhile pointing out that this increase of transverse chromo-electromagnetic fields is not due to the decrease of the effective mass but comes from the decrease of the longitudinal chromo-electromagnetic fields. 

\item[(iii)] For $y_{\rm P}<\eta_{\rm P}$, Eq. \eqref{eq:eta_eom} predicts that the pseudorapidity $\eta_{\rm P}$, like $y_{\rm P}$, also decreases with time. This decrease of $\eta_{\rm P}$ is purely kinematic in nature as a way to respond to any changes of $y_{\rm P}$. On the other hand,  Eq. \eqref{eq:slab_yM_eom} predicts that the rate of momentum space rapidity loss $dy_{\rm P}/d\tau$ becomes smaller and smaller and finally saturates due to the time dependent properties of the functions $\mathcal{A}(\tau)$ and $\mathcal{B}(\tau)$. 
\end{enumerate}

To summarize, the nonvanishing, positive function $\mathcal{B}(\tau)$ in the glasma energy-momentum tensor, which is due to the transverse chromo-electromagnetic fields that gradually build up, is responsible for the nuclear excitation and prevents the momentum space rapidity from forever decreasing. 

\subsection{Determining the Input Parameters}

Initial conditions must be specified. For nucleus-nucleus collisions at center-of-momentum collision energy per nucleon pair $\sqrt{s_{NN}}$, the initial rapidity $y_{\rm P}(\tau=0)$ is computed by
\begin{equation}\label{eq:compute_y0}
y_{\rm P}(\tau=0)\equiv y_0 = \cosh^{-1}\left(\frac{\sqrt{s_{NN}}}{2m_N}\right)
\end{equation}
where $m_N = 0.937 (8) \,\rm{GeV}$ is the nucleon mass. All tubes have the same initial momentum space rapidity. The initial value of pseudorapidity $\eta_{\rm P}(\tau = 0)$ is indeterminate, so we choose the initial value of $\eta_{\rm P}$ to be equal to $y_0$ and start numerical calculations at an infinitesimal initial time $\tau = 0^+$.  Since we consider nucleus-nucleus collisions as a sum of tube-tube collisions at different $\mathbf{r}_{\perp}$, the initial effective mass, which is the mass per unit area, depends on the transverse position as
\begin{equation}\label{eq:initial_mass_scaling}
\mathcal{M}_{\rm P}(\mathbf{r}_{\perp}, \tau=0) = T_A(\mathbf{r}_{\perp})\, m_N. 
\end{equation}
Here $T_A(\mathbf{r}_{\perp}) = \int dz\, \rho_A(\mathbf{r}_{\perp}, z)$ is the nuclear thickness function for a nucleus with mass number $A$. The function $\rho_A(\mathbf{r}_{\perp},z)$ is the Woods-Saxon distribution for a spherical nucleus.

Apart from the initial conditions, other input parameters come from the glasma energy-momentum tensor. Specifically, the functions $\mathcal{A}(\tau)$ and $\mathcal{B}(\tau)$ in Eq. (\ref{eq:glasma_em_tensor}) factorize into two parts: the overall normalization $\varepsilon_0$ which is independent of time, and the time-dependent evolution functions. They can be expressed as
 \begin{equation}
\begin{split}
&\mathcal{A}(\tau) = \varepsilon_0 F_{\mathcal{A}}\left(\ln{(Q^2/m^2)}, Q\tau\right)\, ,\\
&\mathcal{B}(\tau) = \varepsilon_0 F_{\mathcal{B}}\left(\ln{(Q^2/m^2)}, Q\tau\right)\, ,\\
\end{split}
\end{equation}
The functions $F_{\mathcal{A}}$ and $F_{\mathcal{B}}$ depend on two free parameters, the ultraviolet cutoff scale $Q$ and the infrared cutoff scale $m$ on the transverse momentum. The ultraviolet cutoff scale characterizes the division between a description in terms of classical gluon fields and perturbative QCD . Larger values of $Q$ attribute more energy and momentum to the classical fields whereas smaller values of $Q$ attribute more to the production of partons or minijets. The infrared cutoff is identified as the $\Lambda_{\rm QCD}$ scale. The main complication comes from the initial energy density $\varepsilon_0$. To facilitate further discussions, we quote its expression here from Ref. \cite{LK2016}.
\begin{equation}\label{eq:initial_energy_quoted}
\varepsilon_0(\mathbf{r}_{\perp}) = 2\pi\alpha_s^3\frac{N_c}{N_c^2-1}\mu_1(\mathbf{r}_{\perp})\mu_2(\mathbf{r}_{\perp})\ln\left(\frac{Q^2_1}{m^2_1}\right)\ln\left(\frac{Q^2_2}{m^2_2}\right)
\end{equation}
Clearly, for different slab-slab collisions characterized by different values of $\mathbf{r}_{\perp}$, the initial energy densities $\varepsilon_0(\mathbf{r}_{\perp})$ are different. This is due to the width $\mu(\mathbf{r}_{\perp})$ of color charge fluctuations per unit area. Here $\mu_1(\mathbf{r}_{\perp})$ and $\mu_2(\mathbf{r}_{\perp})$ are the color charge fluctuation widths for the two colliding slabs.  Considering a nucleus, we assume that $\mu_A(\mathbf{r}_{\perp})$ is a sum of contributions from all the nucleons at $\mathbf{r}_{\perp}$, that is $\mu_A(\mathbf{r}_{\perp}) = T_A(\mathbf{r}_{\perp}) \mu_{N}$ \cite{Schenke:2012wb,Schenke:2012fw}. The $\mu_N$, which characterizes the gluon saturation for a nucleon, is related to the gluon saturation scale $Q^2_{s,N}$ up to a logarithmic modification \cite{Lappi:2007ku}.  The proton saturation scale $Q^2_{s,N}$ can be extracted from deep inelastic scattering experimental data \cite{Albacete:2014fwa}. With these considerations, we relate the color charge squared per unit area for a tube at transverse position $\mathbf{r}_{\perp}$ to the tube at the central core region of the nucleus $\mathbf{r}_{\perp} = 0$ by
\begin{equation}
\frac{\mu(\mathbf{r}_{\perp})}{\mu(\mathbf{r}_{\perp} =0)} = \frac{T_A(\mathbf{r}_{\perp})}{T_A(\mathbf{r}_{\perp} =0)}. 
\end{equation}
As a consequence, the initial energy density for a tube-tube collision at transverse position $\mathbf{r}_{\perp}$ scales as
\begin{equation}\label{eq:initial_energy_scaling}
\frac{\varepsilon_{0}(\mathbf{r}_{\perp})}{\varepsilon_{0}(\mathbf{r}_{\perp} =0)} = \left[\frac{T_A(\mathbf{r}_{\perp})}{T_A(\mathbf{r}_{\perp} =0)}\right]^2. 
\end{equation}
In arriving at Eq. \eqref{eq:initial_energy_scaling}, we assume the ultraviolet cutoff $Q$ and the infrared cutoff $m$ are the same for all the slabs given the same two colliding nuclei for a fixed collision energy. For asymmetric collisions, the respective nuclear thickness function has to be used. To determine the initial energy density at the central core $\varepsilon_0(\mathbf{r}_{\perp} =0)$, initial conditions for hydrodynamic equations are invoked. Hydrodynamic equations require the initial value of the energy density $\varepsilon_{\rm hydro}(\tau_0)\equiv\varepsilon_{\rm hydro} (x=0,y=0; \tau =\tau_0)$ at spatial location $x=0, y=0$ with $\tau_0$ the moment when hydrodynamics begins.  Both $\varepsilon_{\rm hydro} (\tau_0)$ and $\tau_0$ are free input parameters of hydrodynamic simulations and are tuned to reproduce the experimental data. We assume the initial classical gluon fields from the glasma state are valid until $\tau=\tau_0$ when it is switched to the hydrodynamic state.  This is in the same spirit as the IP-Glasma model \cite{Schenke:2012wb,Schenke:2012fw}. Therefore, the glasma energy density $\varepsilon_{\rm glasma} = \mathcal{A}(\tau) + \mathcal{B}(\tau)$ (see Eq. \eqref{eq:glasma_em_tensor}) at the central core equals $\varepsilon_{\rm hydro} (\tau_0)$  at $\tau=\tau_0$: 
\begin{equation}\label{eq:determine_initial_energy_density}
\varepsilon_0(\mathbf{r}_{\perp} =0) \left[F_{\mathcal{A}}\left(\ln{(Q^2/m^2)}, Q\tau_0\right)+F_{\mathcal{B }}\left(\ln{(Q^2/m^2)}, Q\tau_0\right)\right] = \varepsilon_{\rm hydro}(\tau_0). 
\end{equation}
Hence $\varepsilon_0(\mathbf{r}_{\perp} =0)$ can be solved once $\varepsilon_{\rm hydro}(\tau_0)$ and $\tau_0$ are given. 

\subsection{Numerical Results}

\begin{figure}[th]
 \centering
 \includegraphics[scale=0.85]{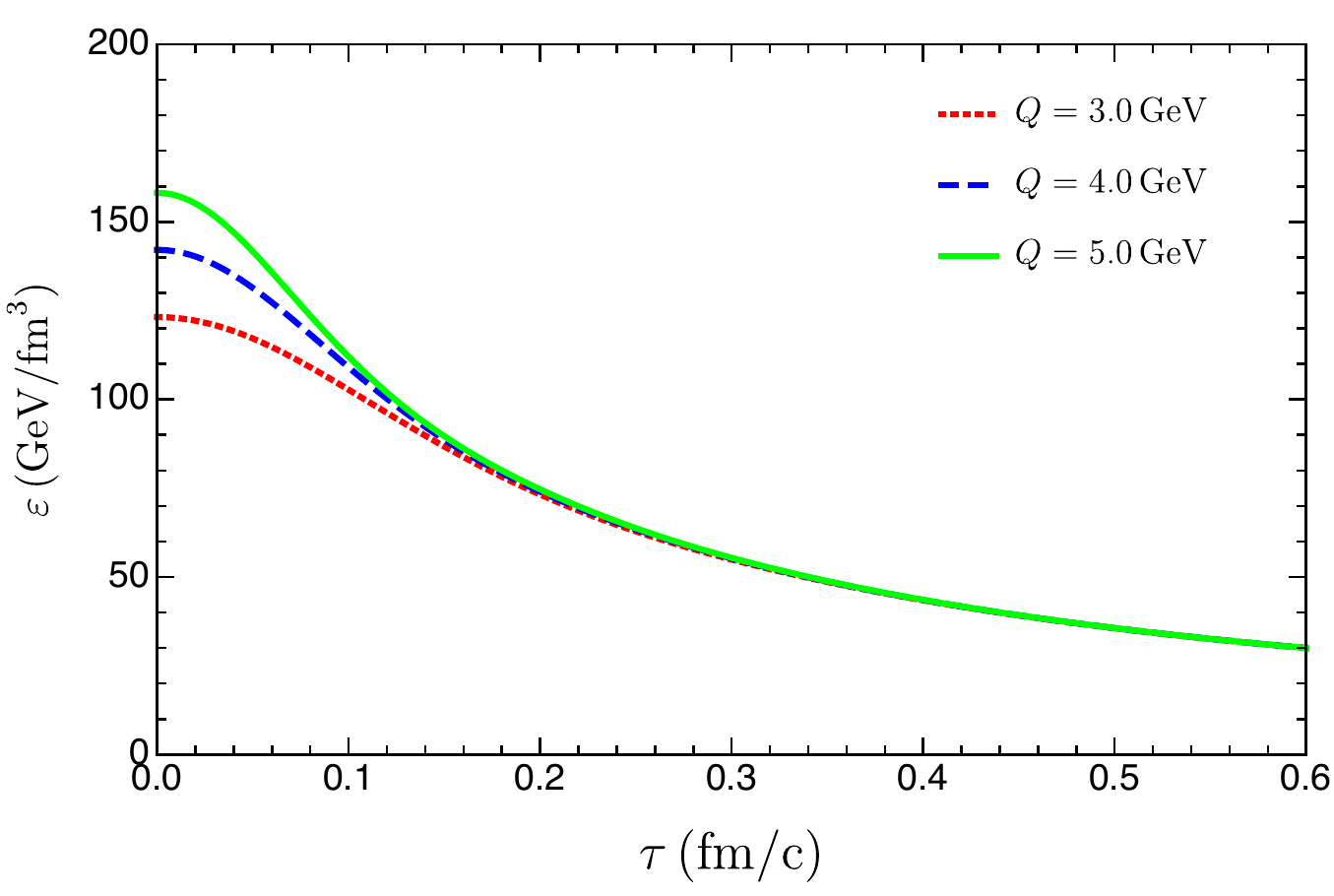}
 \caption{Time evolution of the energy density for different values of the ultraviolet cutoff scale $Q$ given the hydrodynamic initial energy density $\varepsilon_{\rm hydro}(\tau_0 = 0.6\,\rm{fm/c}) = 30.0\,\rm{GeV/fm}^3$.}
 \label{fig:enegy_vs_tau_diff_Q}
\end{figure}

\begin{figure}[th]
 \centering
 \includegraphics[scale=0.80]{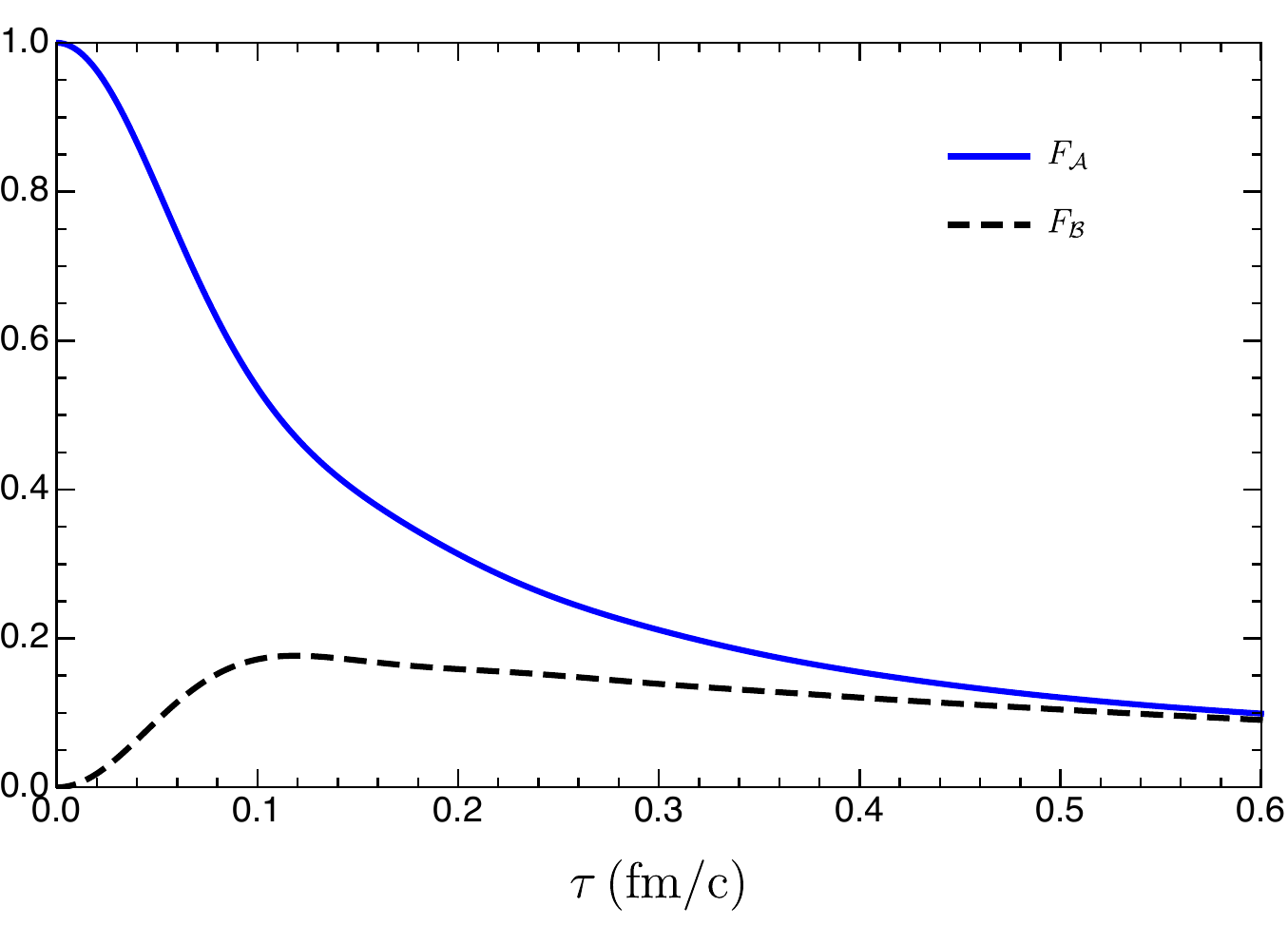}
 \caption{The dependence of $F_{\mathcal{A}} (\tau)$ and $F_{\mathcal{B}}(\tau)$ on proper time for $Q=4.0\,\rm{GeV}$.}
 \label{fig:FA_FB_tau_Q4}
\end{figure}

In this subsection we consider Au+Au collisions at zero impact parameter at the center-of-momentum collision energy $\sqrt{s_{NN}} = 200\,\rm{GeV}$. We focus on the tube-tube collision that comes from the central core of the nucleus characterized by $\mathbf{r}_{\perp} =0$ and compute the rapidity loss and nuclear excitation energy.  The initial beam rapidity is $y_0 = 5.36$. The infrared cutoff is chosen to be $m=\Lambda_{QCD} = 0.2\,\rm{GeV}$. The initial mass per unit area is $\mathcal{M}_{\rm P}(\mathbf{r}_{\perp} =0) = 2.03 \,\rm{GeV/fm}^2$. We use the hydrodynamic initial energy density $\varepsilon_{\rm hydro}(\tau_0 = 0.6\,\rm{fm/c}) = 30.0\,\rm{GeV/fm}^3$, which has been used in viscous hydrodynamic simulations in Ref. \cite{Song:2010aq}. Depending on the ultraviolet cutoff chosen, $\varepsilon_0(\mathbf{r}_{\perp})$ has values $123.2,\, 142.0$ and $158.1\,\rm{GeV/fm}^3$ for $Q = 3.0, \, 4.0$ and $5.0 \, \rm{GeV}$, respectively; see Fig. \ref{fig:enegy_vs_tau_diff_Q}. Different values of the ultraviolet cutoff $Q$ only influence the time evolution of energy density at very early times ($\tau\lesssim 0.15$); all the energy densities converge to the same values at later time when the transition to quark-gluon plasma is assumed to occur. Typical time dependences of $F_{\mathcal{A}}(\tau)$ and $F_{\mathcal{B}}(\tau)$ for $Q=4.0\,\rm{GeV}$ are given in Fig. \ref{fig:FA_FB_tau_Q4}. 

\begin{figure}[h]
 \centering
 \includegraphics[scale=0.85]{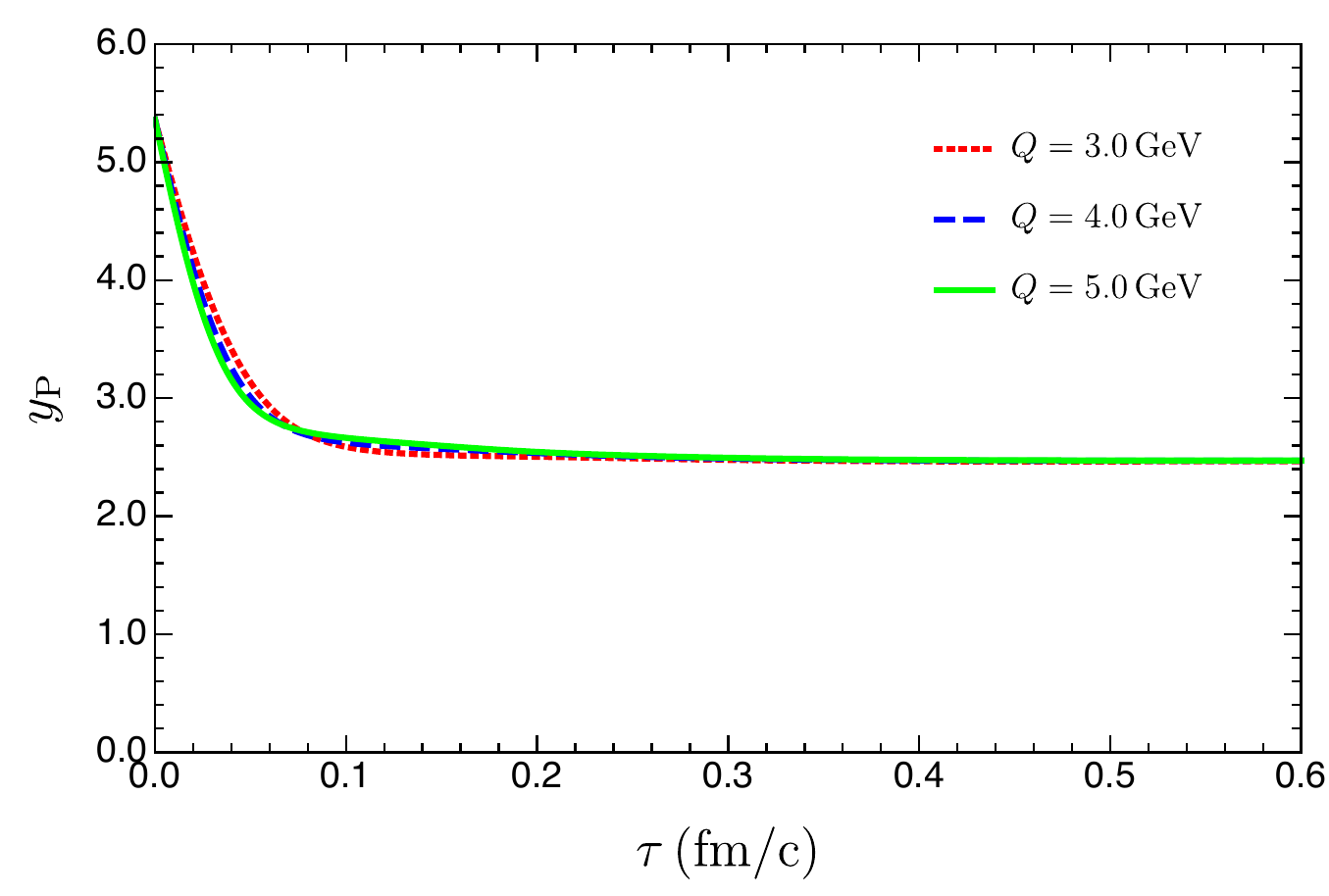}
 \caption{Rapidity of the central core of a gold projectile nucleus in
the center-of-momentum frame for $\sqrt{s_{NN}} = 200\,\rm{GeV}$ as a function of proper time. The result is insensitive to the choice of $Q$ in the physically relevant range.}
 \label{fig:yp_tau_diff_Q}
\end{figure}

With these input parameters we solve the equations of motion \eqref{eq:slab_yM_eom} and \eqref{eq:eta_eom} up to proper time $\tau = 0.6\,\rm{fm/c}$. Figure \ref{fig:yp_tau_diff_Q} shows the momentum-space rapidity $y_{\rm P}$ of the central core of a gold nucleus as a function of proper time $\tau$. The final rapidity in the center-of-momentum frame is 2.47. The central core loses about 2.9 units of rapidity within the first 0.1-0.2 fm/c; this is a robust result, insensitive to the value of $Q$. The rate of rapidity loss decreases with time at the beginning and finally approaches values close to zero, unlike the forever decreasing rapidity in string models. 

\begin{figure}[!pt]
 \centering
 \includegraphics[scale=0.85]{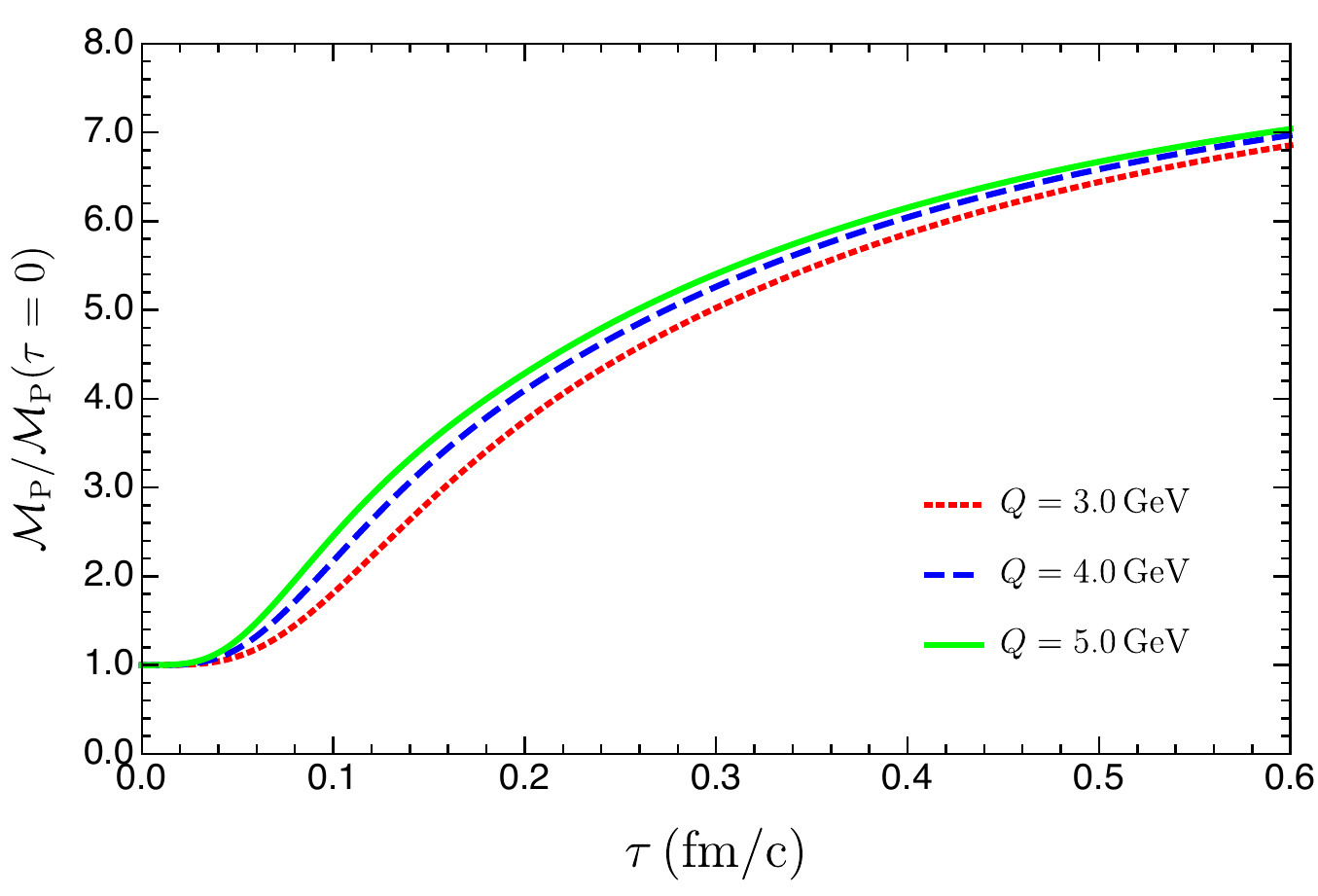}
 \caption{Excitation energy per baryon in the central core of a gold
projectile nucleus in the center-of-momentum frame for $\sqrt{s_{NN}} = 200\,\rm{GeV}$ as a function of proper time. The result is mildly sensitive to the choice of $Q$ in the physically relevant range.}
 \label{fig:mP_tau_diff_Q}
\end{figure}

\begin{figure}[!h]
 \centering
 \includegraphics[scale=0.85]{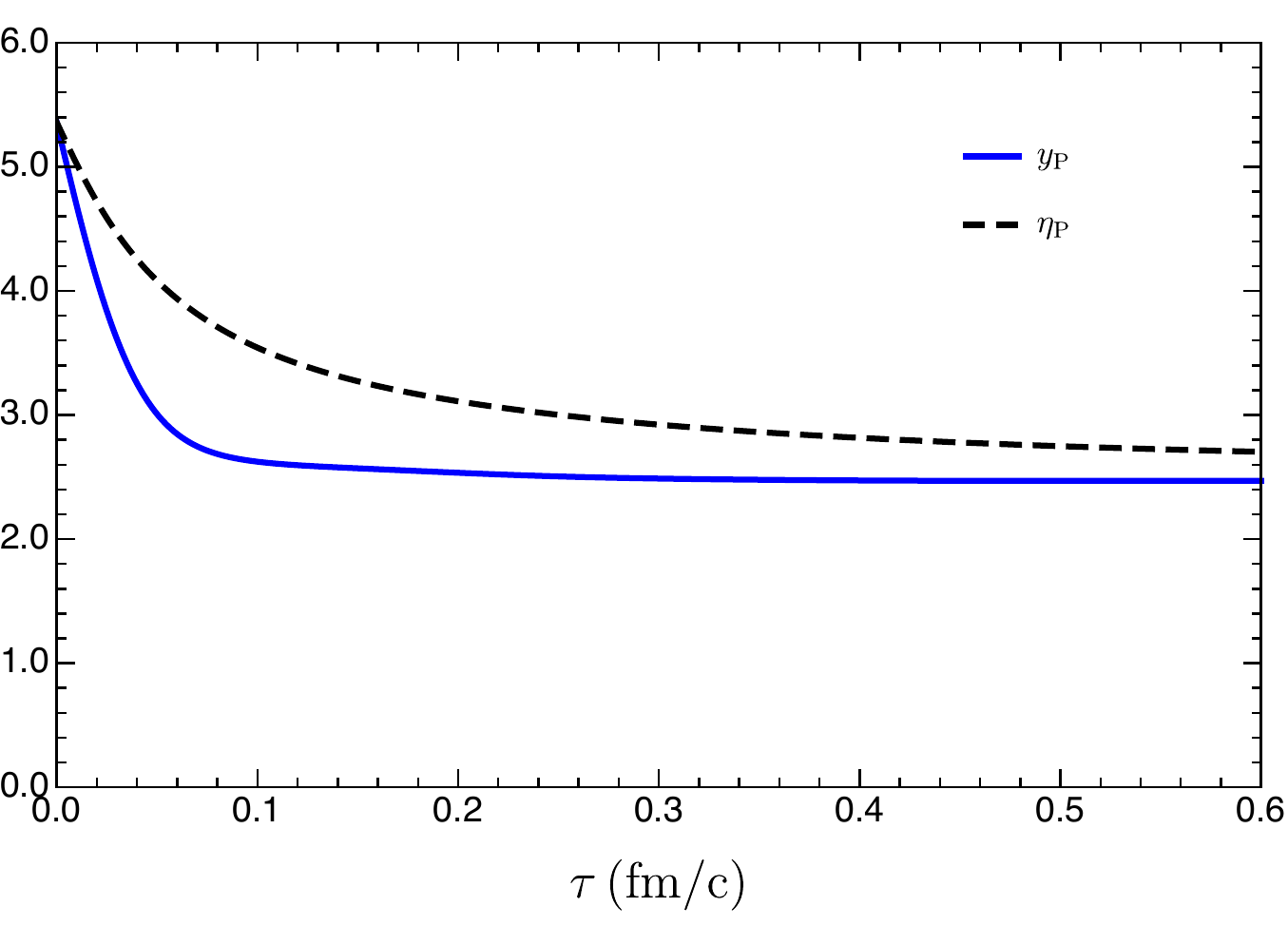}
 \caption{Momentum space rapidity $y_{\rm P}$ and coordinate space pseudorapidity $\eta_{\rm P}$ as functions of proper time for the central core of a gold
projectile nucleus in the center-of-momentum frame for $\sqrt{s_{NN}} = 200\,\rm{GeV}$. The ultraviolet cutoff is $Q=4.0\,\rm{GeV}$. The condition $y_{\rm P} <\eta_{\rm P}$ is maintained in the physically relevant proper time range.}
 \label{fig:yP_etaP_compare_Q4}
\end{figure}

Figure \ref{fig:mP_tau_diff_Q} shows the excitation energy per baryon in units of the nucleon mass as a function of proper time. There is a slow but monotonic increase, unlike the rapidity loss whose asymptotic limit is reached within a few tenths of a fm/c. There is a weak dependence on $Q$. As can be seen from Eq. \eqref{eq:slab_yM_eom}, the increasing rate of the nuclear excitation energy is determined by $2\mathcal{B}(\tau)\sinh{(\eta_{\rm P} - y_{\rm P})}$. At late time, $\mathcal{B}(\tau)$ maintains an almost constant positive value as shown in Fig. \ref{fig:FA_FB_tau_Q4} while the difference $\eta_{\rm P} - y_{\rm P}$ slowly diminishes but maintains finite positive value as shown in Fig. \ref{fig:yP_etaP_compare_Q4}. As a consequence, the increasing rate of the excitation energy gradually decreases.  At $\tau=0.6\,\rm{fm/c}$ the excitation energy reaches $\mathcal{M}_{\rm P}/\mathcal{M}_{\rm P}(\tau=0) = 6.97 $, approximately seven times larger than the nucleon rest mass.  
 

\section{High Baryon and Energy Densities Achievable at Top RHIC Energy}
\label{section: Au_Au200GeV}

In the previous section we calculated the rapidity loss and excitation energy for one single tube-tube collision that comes from the central core of the nucleus. In this section, we repeat the calculations for all the tube-tube collisions at different $\mathbf{r}_{\perp}$ with the help of Eqs. \eqref{eq:initial_mass_scaling} and \eqref{eq:initial_energy_scaling} for central Au+Au collisions at $\sqrt{s_{NN}} = 200\,\rm {GeV}$. Different tube-tube collisions characterized by $\mathbf{r}_{\perp}$ will produce different rapidity losses $y_{\rm P}(\mathbf{r}_{\perp}, \tau)$ and different excitation energies $\mathcal{M}_{\rm P}(\mathbf{r}_{\perp}, \tau)$. Once $y_{\rm P}(\mathbf{r}_{\perp}, \tau)$ and $\mathcal{M}_{\rm P}(\mathbf{r}_{\perp}, \tau)$ have been obtained, the average rapidity loss is readily computed, which is constrained by the experimental data from the BRAHMS collaboration \cite{Bearden:2003hx,  Arsene:2009aa}. Furthermore, rapidity loss and excitation energy predict the baryon density and energy density that can be achieved. Additional thermodynamic properties of the high baryon density matter like temperature and baryon chemical potential can be calculated with the help of an appropriate equation of state. 

\subsection{Net-Baryon Rapidity Distribution}

\begin{figure}[thp]
 \centering
 \includegraphics[scale=1.0]{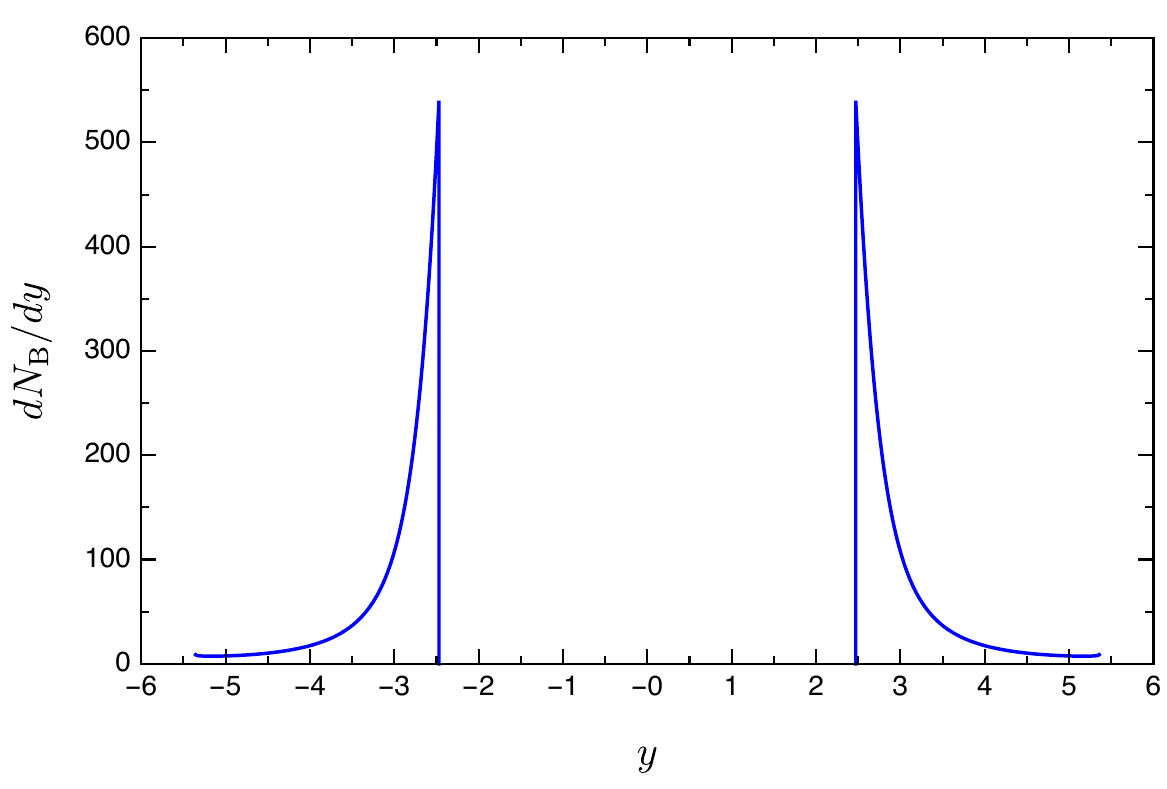}
 \caption{Net-baryon rapidity distribution at $\tau = 0.6\,\rm{fm/c}$ in the center-of-momentum frame after the collision for Au+Au at $\sqrt{s_{NN}} = 200\,\rm{GeV}$. The initial beam rapidities are $y_0 =\pm 5.36$. }
 \label{fig:dNB_dy_vs_y}
\end{figure}

By collecting all the baryons at different $\mathbf{r}_{\perp}$ after the collision at proper time $\tau_0$ and recording their final rapidity $y_{\rm P}(\mathbf{r}_{\perp}, \tau_0)$, one obtains the net-baryon rapidity distribution at that time. Let $y_{\rm P}(\mathbf{r}_{\perp})$ denote the rapidity for the projectile tube at $\mathbf{r}_{\perp}$ at proper time $\tau_0=0.6\,\rm{fm/c}$. The net-baryon rapidity distribution $dN_B/dy$ is computed by summing up all the baryons at different $\mathbf{r}_{\perp}$ that have rapidity $y$.
\begin{equation}\label{eq:netB_dis_no_thermal}
\frac{dN_B}{dy} =\int_0^{2\pi} d\phi \int_{0}^{R_A} r_{\perp} dr_{\perp}\, T_A(\mathbf{r}_{\perp})\,\delta(y-y_{\rm P}(\mathbf{r}_{\perp}))=\frac{2\pi r_{\perp}T_A(r_{\perp})}{|dy_{\rm P}/dr_{\perp}|}\bigg\vert_{r_{\perp} = r_{\perp}(y)}
\end{equation}
Here $R_A$ is the radius of the nucleus and and $T_A(\mathbf{r}_{\perp})$ the nuclear thickness function. We assume azimuthal symmetry in the transverse plane. The total number of baryons should be equal to the nuclear mass number $A=\int_0^{y_0} \frac{dN_B}{dy} dy$ with $y_0$ the initial beam rapidity.  The average rapidity loss follows from
\begin{equation}
\langle \delta y \rangle = y_0 - \frac{1}{A} \int_0^{y_0} y\,\frac{dN_B}{dy} dy. 
\end{equation}
Figure \ref{fig:dNB_dy_vs_y} shows the net-baryon rapidity distribution at $\tau=0.6\,\rm{fm/c}$ after the collision of gold nuclei at $\sqrt{s_{NN}} = 200\,\rm{GeV}$. The initial beam rapidities are $y_0 = \pm 5.36$ and the final rapidities for the central core of the Au nucleus are $y_{\rm P}(\mathbf{r}_{\perp} =0) = \pm 2.47$. The central core of the gold nucleus experiences the largest rapidity loss while the peripheral part ($\mathbf{r}_{\perp} \sim R_A$ ) experiences the smallest rapidity loss. For now we ignore possible thermal motion of baryons inside the nuclear tubes so that all the baryons at $\mathbf{r}_{\perp}$ have the same rapidity $y_{\rm P}(\mathbf{r}_{\perp})$. That is why there is a sharp vertical line at $y=2.47$.  The average rapidity loss is computed to be $\langle \delta y\rangle \approx 2.4$. The BRAHMS collaboration \cite{Bearden:2003hx, Arsene:2009aa} was the only detector at the RHIC that could measure particle production anywhere near the fragmentation regions. The coverage was limited to $|y| \leq3.1$, so the uncertainty in the rapidity loss estimate was large. For 0-5\% centrality BRAHMS found an average rapidity loss of about $2.05+0.4/−0.6$. This is consistent with our result, especially since we focus on $0\%$ centrality for illustration.

\newpage  

\subsection{Large Baryon and Energy Densities}

With the rapidity loss $y_{\rm P}(\mathbf{r}_{\perp},\tau)$ and the excitation energy $\mathcal{M}_{\rm P}(\mathbf{r}_{\perp},\tau)$, we are able to calculate the baryon density and the energy density.  The baryon density at proper time $\tau_0$ is computed by \cite{AKM1980,C1984,GC1986,LKRapid}
\begin{equation}\label{eq:nB_LRF}
n_B(r_{\perp}, z^{\prime}, \tau_0) = e^{\Delta y(r_{\perp},\tau_0)}\, \rho_A(r_{\perp}, z^{\prime}\,e^{\Delta y(r_{\perp},\tau_0)}),
\end{equation}
where $z^{\prime} = z - z_{\rm P}(r_{\perp},\tau_0)$ is the longitudinal coordinate in the local rest frame of the tube characterized by $r_{\perp}$. The change of rapidity $\Delta y(r_{\perp},\tau_0) = y_0 - y_{\rm P}(r_{\perp},\tau_0)$ depends on the transverse position $r_{\perp}$. Different parts of the colliding nucleus characterized by $r_{\perp}$ have different time dependent rapidity $y_{\rm P}(r_{\perp}, \tau)$ and different longitudinal trajectory $z_P(r_{\perp},\tau)$ as viewed in the center-of-momentum frame. Those tubes that are close to the central core of the nucleus experience larger rapidity loss and travel a smaller longitudinal distance $z_{\rm P}(r_{\perp}\sim 0,\tau_0)$ at time $\tau_0$, while those tubes close to the peripheral region of the nucleus experience small rapidity loss and travel to a larger longitudinal distance $z_{\rm P}(r_{\perp}\sim R_A,\tau_0)$. Therefore, different tubes characterized by different $r_{\perp}$ are separated along the longitudinal direction due to different $z_{\rm P}(r_{\perp}, \tau_0)$ at $\tau_0$, and the spherical shape of the nucleus before the collision will no longer be maintained after the collision.  Since $y_{\rm P}(r_{\perp},\tau)$ also depends on $r_{\perp}$, there is no single reference frame that is the local rest frame for all the tubes comprising the fireball.  Each tube has its own local rest frame by boosting the center-of-momentum frame to the frame moving at rapidity $y_{\rm P}(r_{\perp}, \tau_0)$. Multiplying the baryon density by the nuclear excitation energy, one obtains the energy density
\begin{equation}\label{eq:eB_LRF}
\varepsilon(r_{\perp}, z^{\prime},\tau_0) = \frac{\mathcal{M}_{\rm P}(r_{\perp}, \tau_0)}{\mathcal{M}_{\rm P}(r_{\perp}, \tau = 0)}\, m_N \, n_B(r_{\perp}, z^{\prime},\tau_0). 
\end{equation}
The energy density relies on the excitation energy which slowly increases with time, see Fig. \ref{fig:mP_tau_diff_Q}. Hence, the energy density depends on the proper time chosen to evaluate its value. 

\begin{figure}[thp]
 \centering
 \includegraphics[scale=1.0]{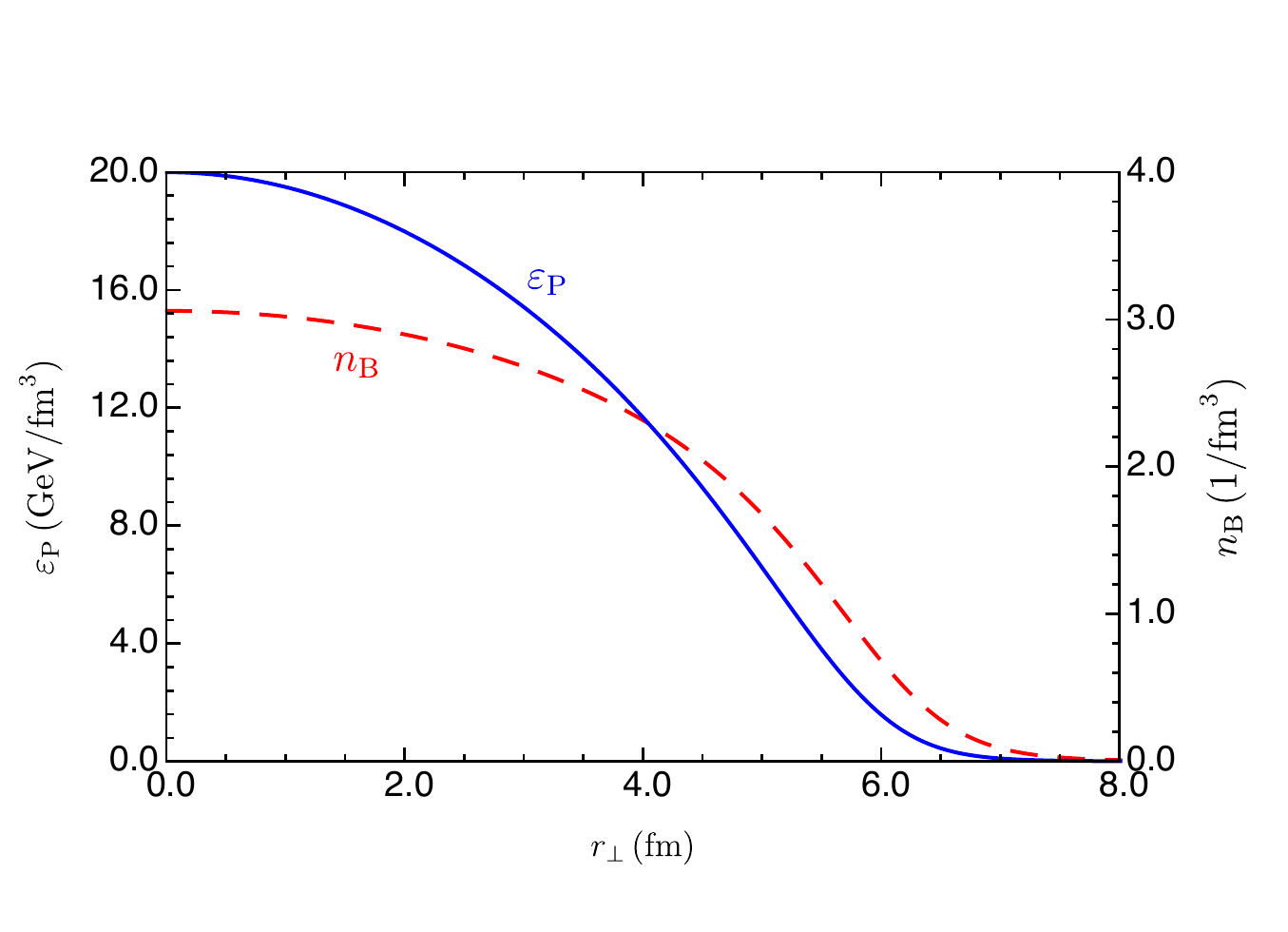}
 \caption{The energy and baryon densities at $\tau = 0.6\, \rm{fm/c}$
as functions of the transverse distance for central collisions of Au nuclei at $\sqrt{s_{NN}}=200\,\rm{GeV}$. }
 \label{fig:eB_nB_rT_AuAu200}
\end{figure}

Figure \ref{fig:eB_nB_rT_AuAu200} shows the proper energy and baryon densities as functions of the transverse coordinate at $\tau = 0.6\,\rm{fm/c}$ for $z^{\prime} =0$. It should be noted that the maximum baryon density, about 3 baryons/fm$^3$, is 20 times greater than the normal matter density of 0.155 nucleons/fm$^3$. The maximum energy density is about $20\,\rm{GeV/fm}^3$, much larger than the critical energy density $\sim1.0\,\rm{GeV/fm}^3$ for the formation of quark-gluon plasma.  

\begin{figure}[thp]
 \centering
 \includegraphics[scale=1.0]{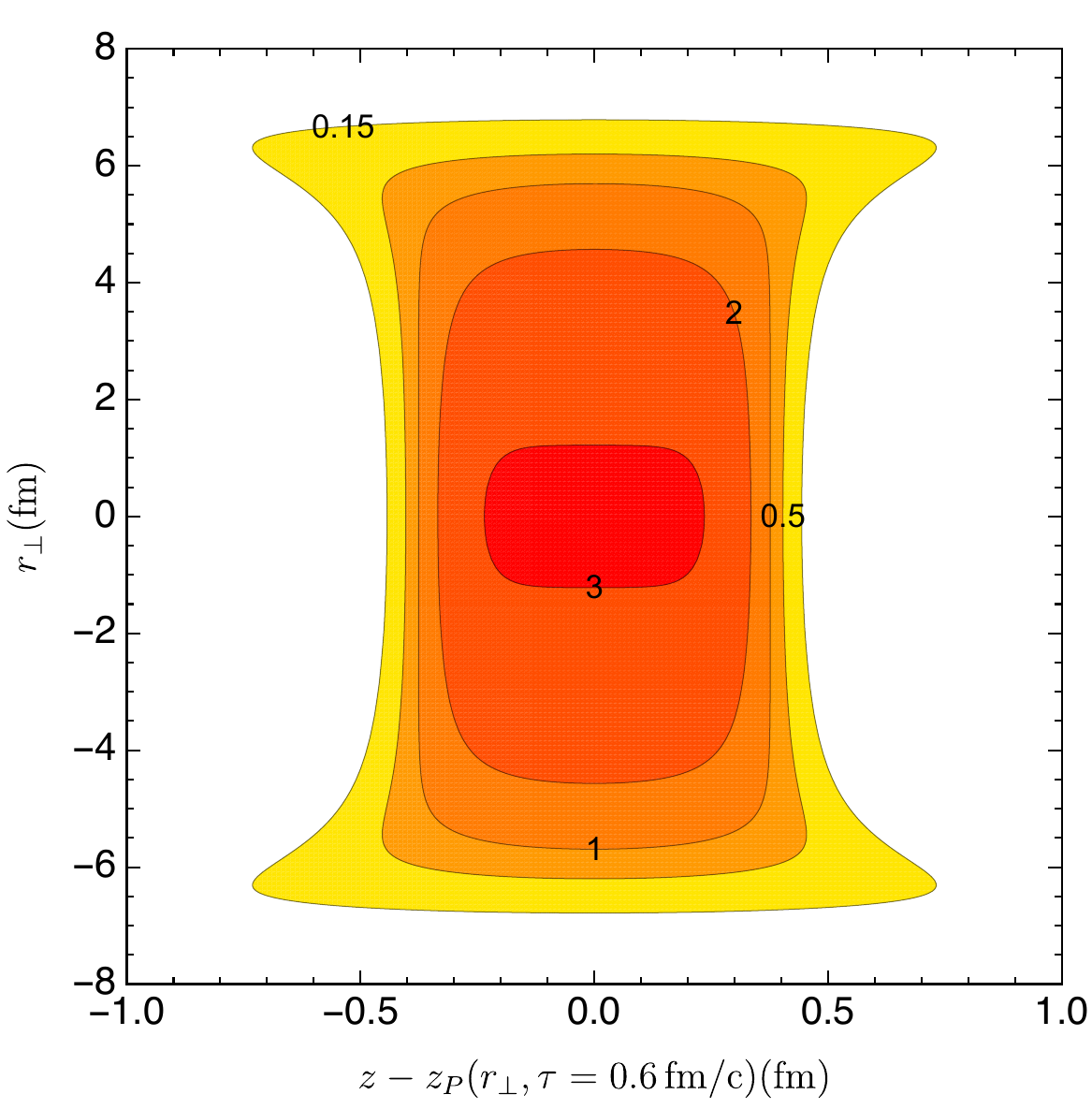}
 \caption{Contour plot of the proper baryon density for central
collisions of Au nuclei at $\sqrt{s_{NN}} = 200\, \rm{GeV}$. The numbers are in units of baryons per fm$^3$. The horizontal axis measures the distance along the beam direction in the local rest frame. Care must be taken when interpreting this plot since the rapidity of the matter and therefore the frame of reference depend on $r_{\perp}$. }
 \label{fig:rT_z_nB_AuAu200}
\end{figure}

Figure \ref{fig:rT_z_nB_AuAu200} is a contour plot of the proper baryon density. The contours are drawn at $n_{\rm B}$ = 3, 2, 1, 0.5, and 0.15 baryons/fm$^3$. The shapes of the contours arise for the following reasons. The diameter of a gold nucleus 2$R_A$ is about 14 fm. The core centered at $r_{\perp}$ = 0 along the z axis contains the most matter, suffers the greatest deceleration, and hence the greatest compression. Moving outward with increasing $r_{\perp}$, the length of the tube is decreased to $2\sqrt{R_A^2 - r_{\perp}^2}$, and
the deceleration, and hence compression, is reduced. These opposing effects approximately cancel each other, giving rise to roughly rectangular contours in the $r_{\perp}$-$z^{\prime}$ plane. Care must be taken when interpreting this figure. Since the rapidity loss depends on $r_{\perp}$ it means that there is a shear in the $r_{\perp}$ direction, and there is no single global frame of reference for all elements of the fireball. It should be emphasized that the baryon densities calculated here are more robust than the energy densities. The reason can be seen by comparing Eqs. \ref{fig:yp_tau_diff_Q} and \ref{fig:mP_tau_diff_Q}. The rapidity loss, and therefore compression, is determined mostly within the first few tenths of a fm/c when the glasma dominates the dynamics. The excitation energy continues its slow growth as time goes on. If the transition from glasma to quark-gluon plasma happens earlier than 0.6 fm/c, it would reduce the excitation energy but hardly affect the compression. Exactly how the transition occurs is a topic of much current interest and activity. This should be kept in mind in the following discussions.

\newpage  

\begin{figure}[tp]
 \centering
 \includegraphics[scale=0.99]{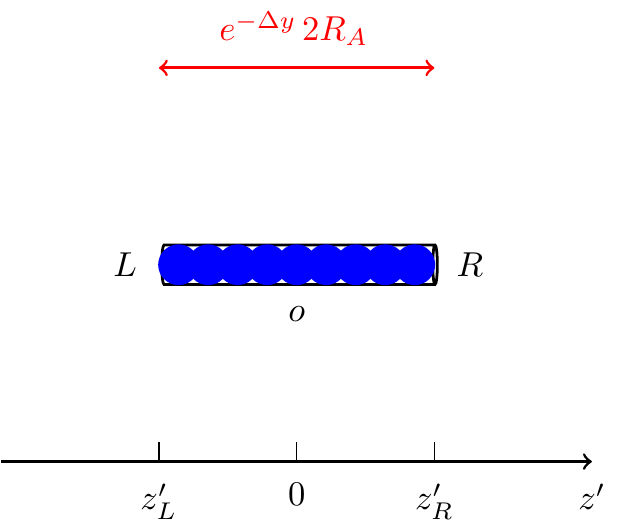}
 \caption{Schematic picture of the tube for the central core of a nucleus in its local rest frame.}
 \label{fig:nuclear_slab_LRF}
\end{figure}

Both the baryon density Eq. \eqref{eq:nB_LRF} and the energy density Eq. \eqref{eq:eB_LRF} are evaluated in the local rest frame of the tube characterized by $r_{\perp}$. In this representation, Fig. \ref{fig:rT_z_nB_AuAu200} displays the volume information of the fireball. To present the distributions in the center-of-momentum frame, where the spatial distribution of baryons is apparent, requires making Lorentz boosts from the individual local rest frame characterized by $r_{\perp}$ to the center-of-momentum frame. Note that the baryons shown in Fig. \ref{fig:rT_z_nB_AuAu200} share the same proper time $\tau$ instead of the same Cartesian coordinate time $t$. Figure \ref{fig:nuclear_slab_LRF} is a schematic picture of the central tube after collision in its own rest frame. Let ($t^{\ast}, z^{\ast}$) denote Cartesian coordinates in the center-of-momentum frame and ($t^{\prime}, z^{\prime}$) denote Cartesian coordinates in the local rest frame of the nuclear slab. Making the Lorentz transformation one gets
\begin{equation}
z^{\prime}_L = \gamma_{\rm P} (z^{\ast}_L - \beta_{\rm P} t^{\ast}_L),
\end{equation}
where $\gamma_{\rm P} = \cosh{y_{\rm P}}$ and $\beta_{\rm P} = \tanh{y_{\rm P}}$ with $y_{\rm P}$ the rapidity of the tube. The baryons all have the same proper time $\tau$ when viewed in the center-of-momentum frame so that $z^{\ast}_L = \tau \sinh{\eta_L}$ and $t^{\ast}_L = \tau \cosh{\eta_L}$. Here $\eta_L$ is the pseudorapidity defined in the center-of-momentum frame. Then $z^{\prime}_L = \tau \sinh{(\eta_L -y_{\rm P})}$
which, compared with $z^{\ast} = \tau \cosh{\eta_L}$, is just a shift of rapidity from $\eta_L$ to $\eta_L-y_{\rm P}$. Therefore, the space-time pseudorapidity $\eta_L$ of a baryon labeled by $L$ in the center-of-momentum frame is related the the corresponding coordinate $z^{\prime}_L$ in the local rest frame by
\begin{equation}\label{eq:z_eta_transformation}
\eta_L = \sinh^{-1}\left(\frac{z^{\prime}_L}{\tau} \right)+ y_{\rm P}\, .
\end{equation}
A potential problem with Eq. \eqref{eq:z_eta_transformation} is that for $z^{\prime}_{\rm O} =0 $ it predicts $\eta_{\rm O} = y_{\rm P}$, which is not exactly true since $\eta_{\rm O}$ slightly deviates from $y_{\rm P}$; see Fig. \ref{fig:yP_etaP_compare_Q4}.  Here O indicates the center of the tube in Fig. \ref{fig:nuclear_slab_LRF}. However, this only slightly influences the absolute position of the pseudorapidity for the center of the tube; the span of the pseudorapidity remains unchanged.  This analysis improves upon and supercedes that reported in Ref. \cite{LKRapid}.

\begin{figure}[thp]
 \centering
 \includegraphics[scale=1.0]{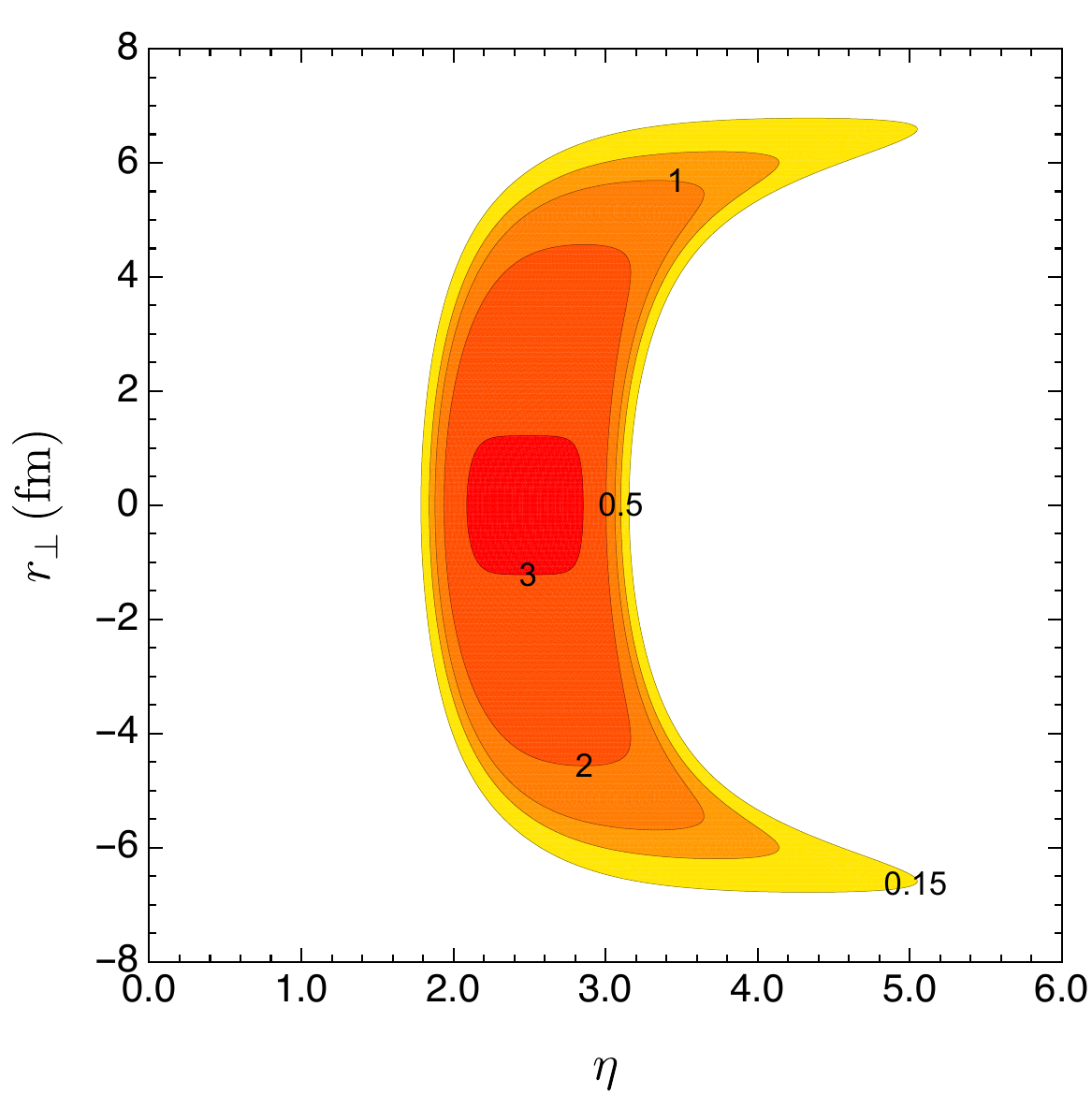}
 \caption{Contour plot of the proper baryon density for central
collisions of gold nuclei at $\sqrt{s_{NN}} = 200\, \rm{GeV}$. The numbers are in units of baryons per fm$^3$. The horizontal axis is the space-time pseudorapidity in the center-of-momentum frame.}
 \label{fig:rT_eta_nB_AuAu200}
\end{figure}

Figure \ref{fig:rT_eta_nB_AuAu200} shows the baryon distribution in the $r_{\perp}$-$\eta$ plane using Eq. \eqref{eq:z_eta_transformation}. The central tube spans about 1.5 units of rapidity. This distribution is useful as an initial condition for the subsequent hydrodynamic evolution in space and time, which is outside the scope of this paper.

\subsection{Temperature and Baryon Chemical Potential}

The previous results and discussions have not assumed the fireballs in the fragmentation regions to be thermalized.  High energy densities are only necessary but not sufficient conditions for the formation of quark-gluon plasma. If and when thermalization occurs is hard to know.  The proper time $\tau=0.6\,\rm{fm/c}$ is the typical time when quark-gluon plasma is believed to be formed in the central region of Au+Au collisions at $\sqrt{s_{NN}} = 200\,\rm{GeV}$ and is also the time when the matter in the central region is assumed to be thermalized.  Exactly how the quark-gluon plasma in the central region equilibrates and thermalizes has not reached a definitive conclusion yet and is still under active investigation \cite{Fukushima:2016xgg}.  The practical approach is to assume the thermalization has been reached and tune the input parameters at $\tau=0.6\,\rm{fm/c}$ (for example) so as to reproduce experimental data. The exact mechanism for the thermalization is actually not so relevant as long as it predicts the required initial conditions at $\tau = 0.6\,\rm{fm/c}$ constrained by experimental data. As for the fragmentation regions, Anishetty, Koehler and McLerran  \cite{AKM1980} argued that the matter in the fireballs in the fragmentation regions could thermalize due to interactions of secondary particles. Since then, there has been very little research concerning the  fragmentation regions, not to mention the thermalization problem. Thermalization in the fragmentation regions, if it is possible, could be as challenging as the thermalization problem in the central region and beyond the scope of this paper. Just like the current practice in the central region, we assume the fireballs have reached thermalization and explore the consequences. 

Assuming local thermal equilibrium, what are the temperature and baryon chemical potential? To answer that question we need an equation of state. In the limit of very high temperature and baryon density, the equation of state can be calculated using perturbative QCD in thermal field theory \cite{Kapusta:2006pm, Kurkela:2016was}. On the other hand, in the low temperature and small baryon density regime, the relevant degrees of freedom are hadrons. For a thermalized system of hadrons, the hadron resonance gas model gives a very good description of the thermodynamic properties of the system.  Furthermore, first principle calculations based on lattice QCD provides robust results for the equation of state of a system of quarks and gluons in a very wide range of temperature for zero baryon chemical potential \cite{Aoki:2006we, Borsanyi:2010cj}. Extending to finite baryon chemical potential has the notorious sign problem. Currently, much effort has been devoted to extending the lattice calculations to finite baryon chemical potential \cite{Bazavov:2017dus}. In the following, we will use a crossover equation of state \cite{Albright} that smoothly connects the quark-gluon plasma phase and the hadronic resonance gas phase consistent with lattice data \cite{Borsanyi:2010cj, Borsanyi:2012cr}.  This equation of state does not contain a first order phase transition line or critical point. Instead, the transition from the quark-gluon plasma phase to the hadronic resonance gas phase is a rapid, smooth crossover both for zero baryon chemical potential and for nonzero baryon chemical potentials. The crossover equation of state \cite{Albright} has the form 
\begin{equation}\label{eq:crossover_EOS_I}
P(T,\mu) = S(T,\mu) P_{\rm{qg}}(T,\mu) + \left[ 1- S(T,\mu)\right] P_{\rm{h}}(T,\mu). 
\end{equation}
with the switching function
\begin{equation}\label{eq:crossover_EOS_II}
S(T,\mu) = \rm{exp}\{ - \theta(T,\mu)\},\quad \theta(T,\mu) = \left[\left(\frac{T}{T_0}\right)^r + \left(\frac{\mu}{\mu_0}\right)^r\right]^{-1}.
\end{equation}
Here $P_{\rm qg}$ represents the perturbative QCD results for the pressure of the quark-gluon plasma phase while $P_{\rm h}$ represents pressure from the excluded volume model of the hadronic resonance gas phase. The switching function $S(T,\mu)$ asymptotically approaches $1$ for very large $T$ and $\mu$ and asymptotically approaches $0$ for very small $T$ and $\mu$. The free parameters $T_0$, $\mu_0$ and $r$ are optimized to be consistent with lattice data. 

\begin{figure}[h]
 \centering
 \includegraphics[scale=1.0]{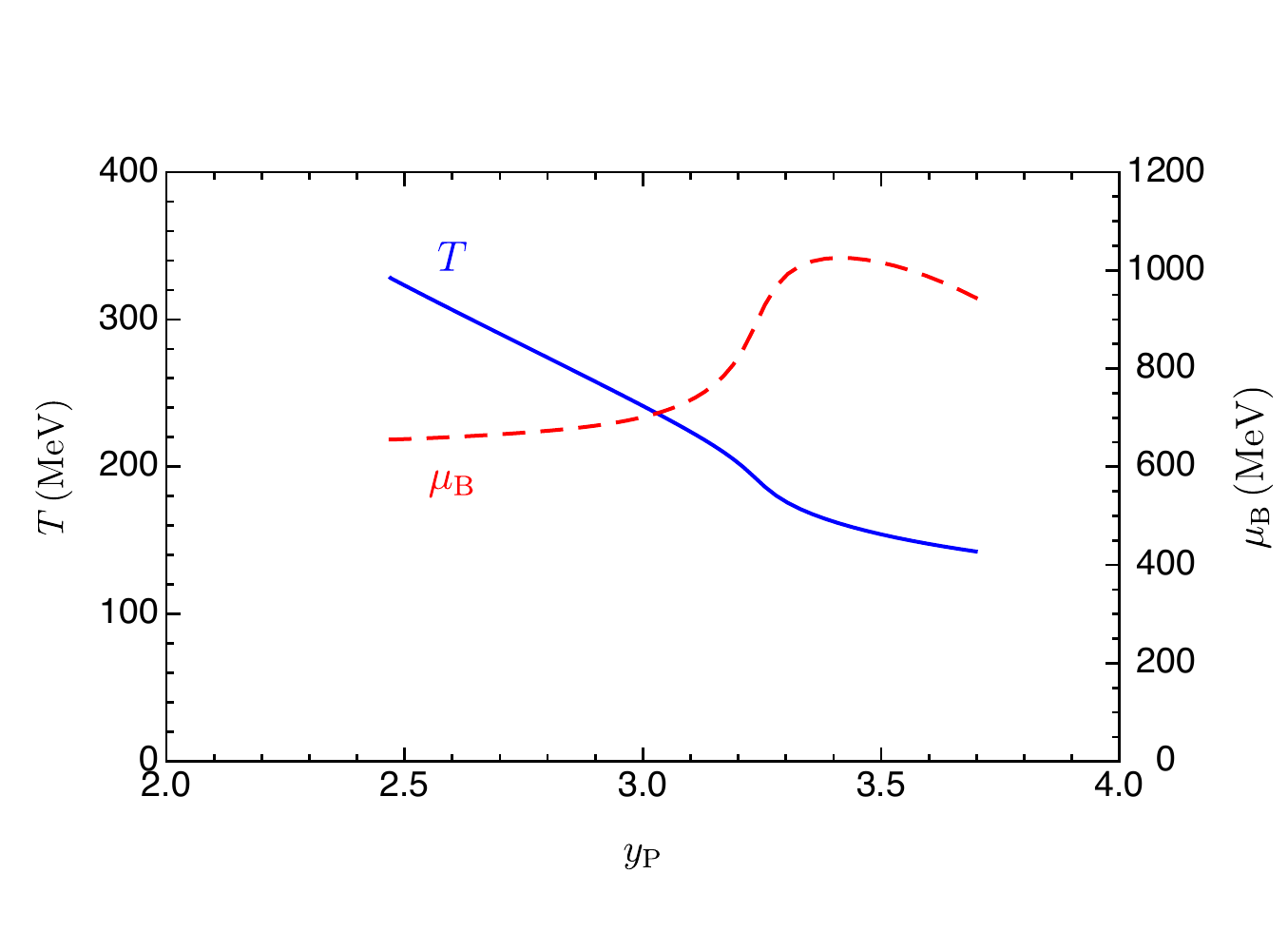}
 \caption{The temperature and baryon chemical potential in the fragmentation region as a function of momentum space rapidity at $\tau=0.6\,\rm{fm/c}$. Values are calculated for $z^{\prime}=0$ as defined in Eq. \eqref{eq:nB_LRF}. Only matter with energy densities larger than $1.0\,\rm {GeV/fm}^3$ have been displayed.}
 \label{fig:temp_muB_vs_y_initial}
\end{figure}

\begin{figure}[h]
 \centering
 \includegraphics[scale=1.0]{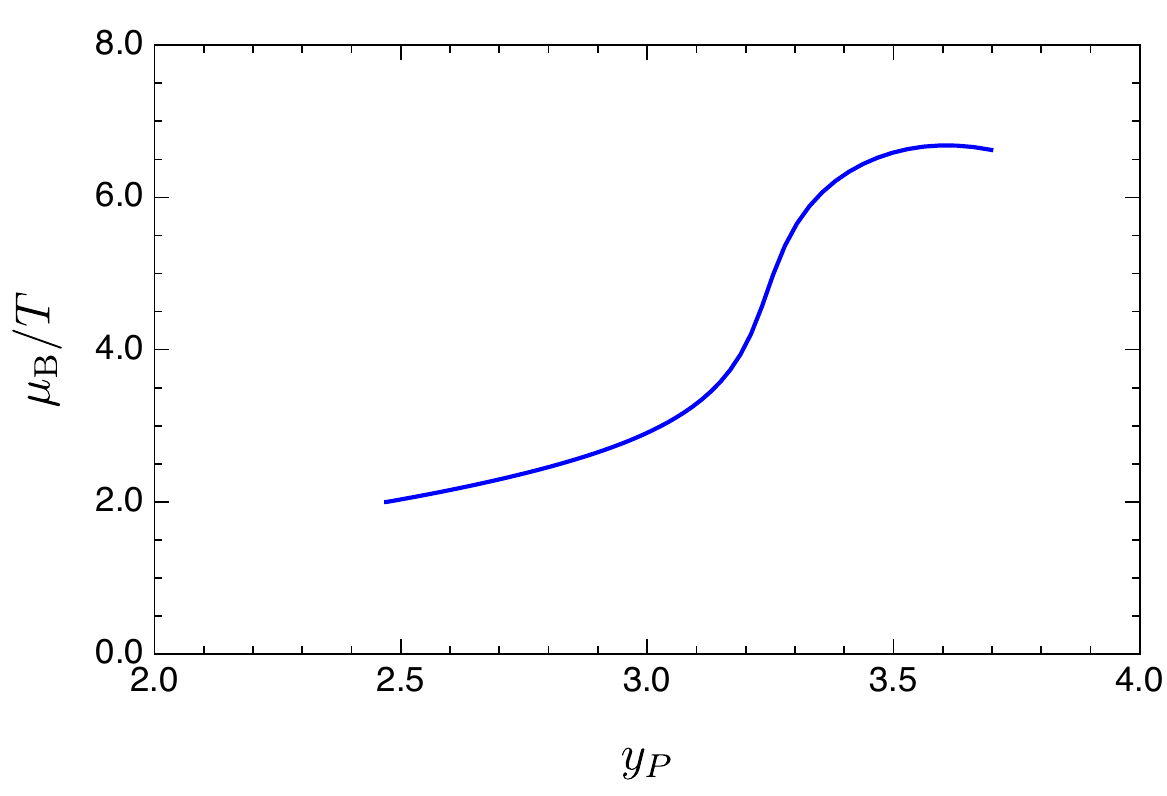}
 \caption{Ratio of baryon chemical potential to temperature as a function of momentum space rapidity at $\tau=0.6\,\rm{fm/c}$.}
 \label{fig:muBoT_vs_yP} 
\end{figure}

\begin{figure}[h]
 \centering
 \includegraphics[scale=1.0]{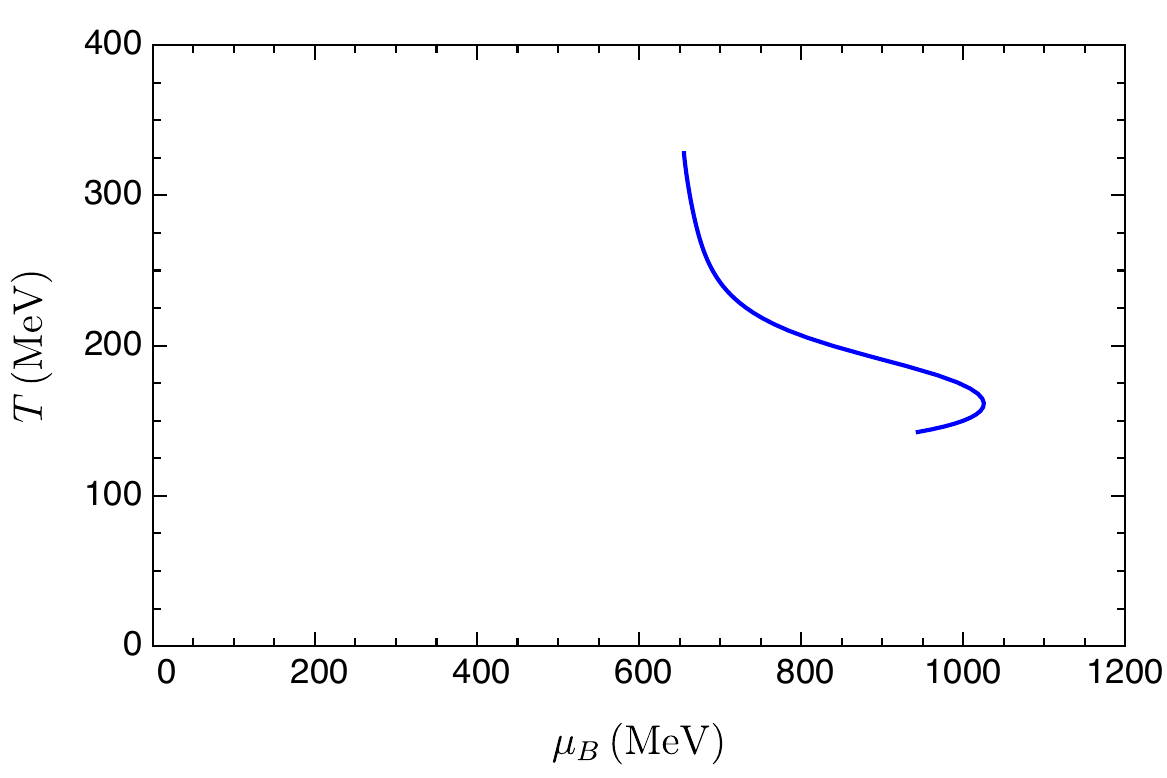}
 \caption{The initial temperatures and baryon chemical potentials in the fragmentation regions at $\tau= 0.6\,\rm{fm/c}$ with energy density larger than $1.0\,\rm{GeV/fm}^3$. Values are calculated for $z^{\prime}=0$ as defined in Eq. \eqref{eq:nB_LRF}.}
 \label{fig:temp_vs_muB_initial} 
\end{figure}

\begin{figure}[h]
 \centering
 \includegraphics[scale=0.9]{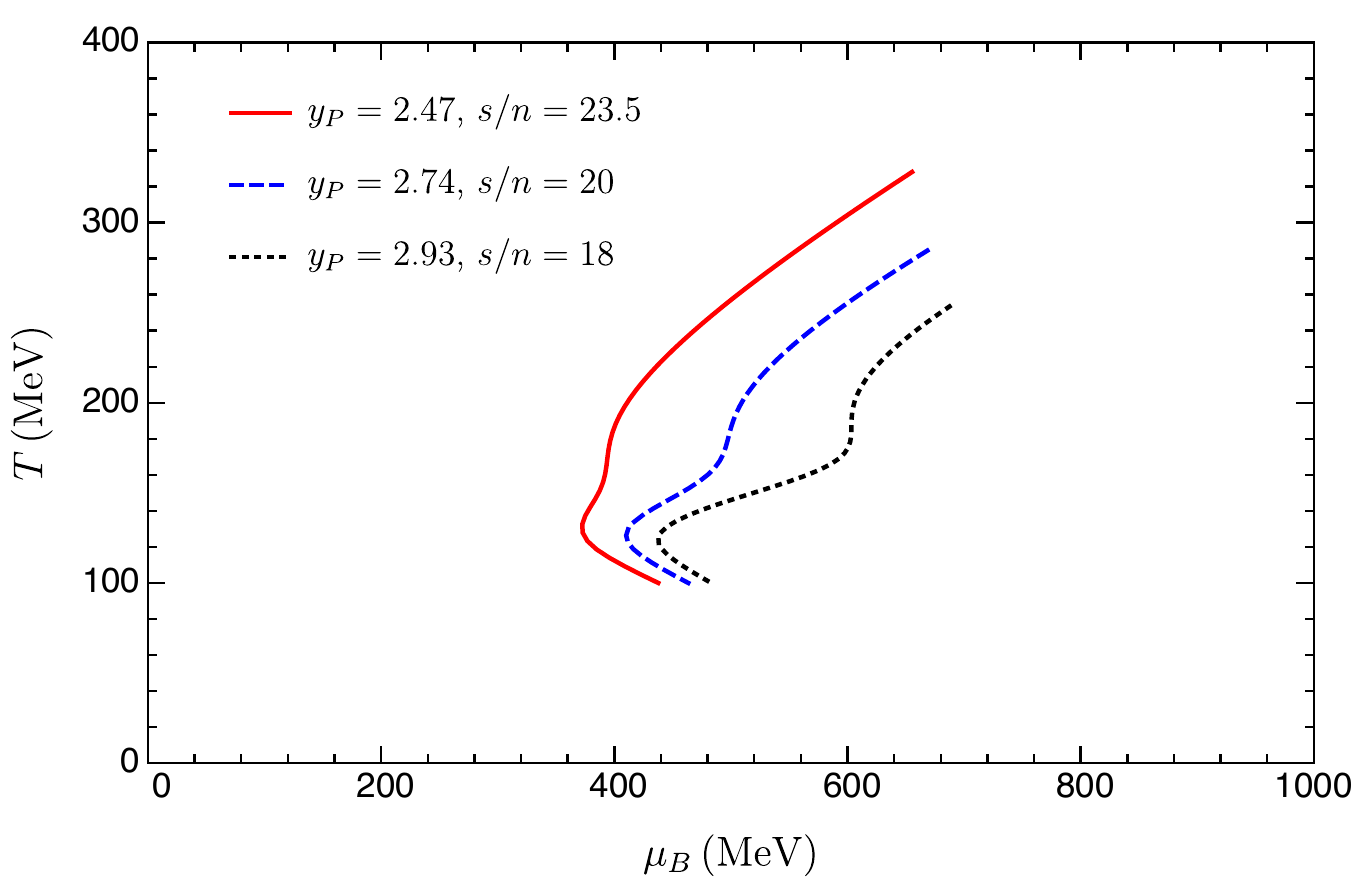}
 \caption{Phase trajectories of adiabatic expansion at three different rapidities and entropy per baryon ratios.}
 \label{fig:adiabatic_phase_trajectories} 
\end{figure}

\begin{figure}[thp]
 \centering
 \includegraphics[scale=1.0]{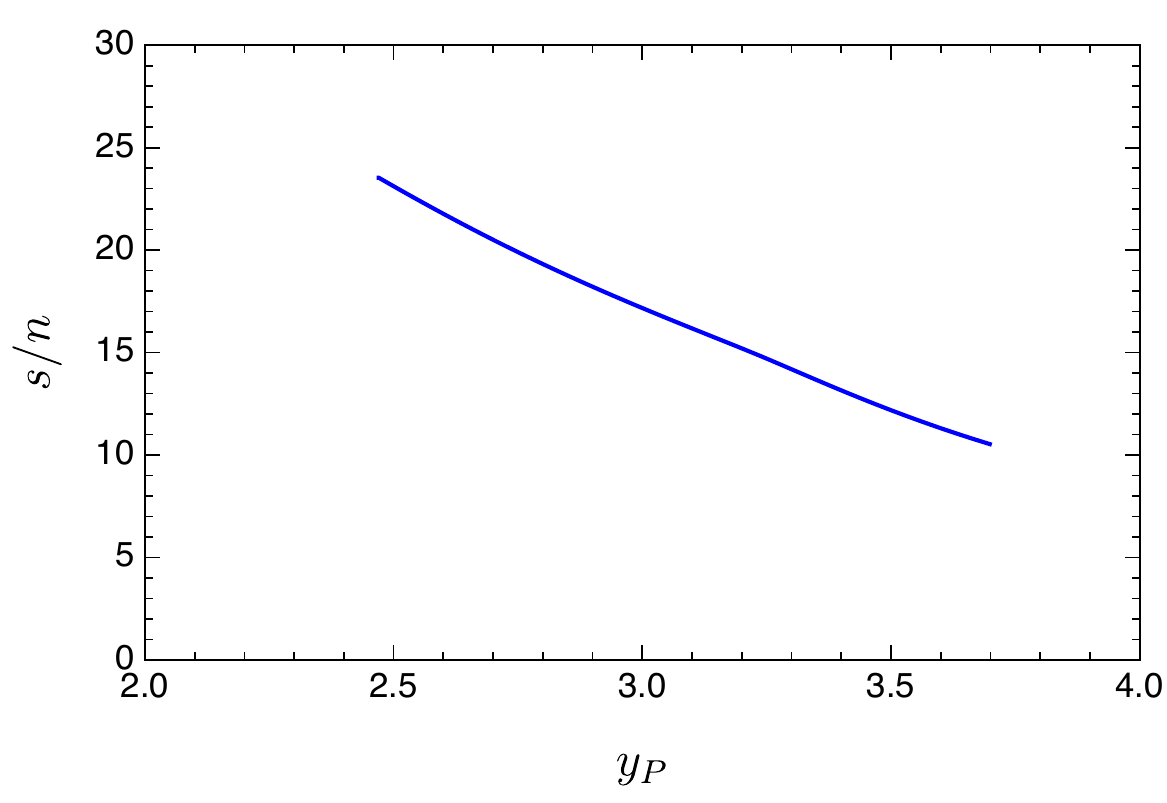}
 \caption{Entropy per baryon as a function of the momentum space rapidity. A rapidity scan may help locate the critical point.}
 \label{fig:son_vs_yP} 
\end{figure}

Given the energy and baryon density distributions for $z^{\prime}=0$ shown in Fig. \ref{fig:eB_nB_rT_AuAu200}, we compute the corresponding temperature and baryon chemical potential distributions using the crossover equation of state from Eqs. \eqref{eq:crossover_EOS_I} and \eqref{eq:crossover_EOS_II}.  Figure \ref{fig:temp_muB_vs_y_initial} shows the temperatures and baryon chemical potentials for different rapidities $y_{\rm P}$ instead of $r_{\perp}$. 
Note that there is a one-to-one correspondence between $r_{\perp}$ and $y_{\rm P}$: $y_{\rm P} = y_{\rm P}(r_{\perp})$. Only those areas with energy densities larger than $1.0\,\rm{GeV/fm}^3$ have been displayed. In the range of rapidity from $2.47$ to $3.45$, the value of the baryon chemical potential increases from $655\,\rm{MeV}$ to $1020\,\rm{MeV}$ while the value of the temperature correspondingly decreases from $328\,\rm{MeV}$ to $155\,\rm{MeV}$. The baryon chemical potential to temperature ratio ranges from $2.0$ to $6.5$ as shown in Fig. \ref{fig:muBoT_vs_yP}. The same data is shown in the $\mu_B$-$T$ plane in Fig. \ref{fig:temp_vs_muB_initial}.  
Assuming the subsequent expansion of the thermalized high baryon density matter after $\tau=0.6\,\rm{fm/c}$ is approximately adiabatic, just like in the central region, then typical phase trajectories in the $\mu_B$-$T$ plane may be computed.  See Fig. \ref{fig:adiabatic_phase_trajectories} for three different entropy per baryon ratios. The corresponding momentum space rapidities are correlated with the entropy per baryon. These phase trajectories tilt to the right after changing from the quark-gluon plasma phase to the hadronic gas phase. In the transition region, the temperature decreases very quickly while the baryon chemical potential decreases very slowly.  From these phase trajectories, it is possible that the expansion of the high baryon density matter in the fragmentation regions might go through or near the region in the $\mu_B$-$T$ plane where a first order phase transition line or a critical point are conjectured to occur \cite{Nonaka:2004pg,Asakawa:2008ti}. As shown in Fig. \ref{fig:son_vs_yP}, the entropy per baryon ratio might be in the right range so that a scan through the momentum space rapidity may locate the critical point.

\section{High Baryon Density Achievable in Other Collision Configurations}
\label{section:systematic_studies}

To extend the previous analysis of the high baryon densities achieved outside the central rapidity region of Au+Au collisions at $\sqrt{s_{NN}} = 200\,\rm{GeV}$ to other heavy ion collision configurations involving different nuclear sizes, different collision energies or different impact parameters, the crucial step is to determine the initial energy density of the glasma $\varepsilon_0(\mathbf{r}_{\perp})$ for these different collision configurations. In the case of central Au+Au collisions at $\sqrt{s_{NN}} = 200\,\rm{GeV}$, the $\varepsilon_0(\mathbf{r}_{\perp})$ is determined by $\varepsilon_0(\mathbf{r}_{\perp} =0)$ through the nuclear thickness function $T_A(\mathbf{r}_{\perp})$, while $\varepsilon_0(\mathbf{r}_{\perp}=0)$ is fixed by the initial energy density of hydrodynamics at $\tau_0=0.6\,\rm{fm/c}$,  assuming that the classical gluon fields are valid up to the time when hydrodynamic evolution of the quark-gluon plasma begins. In principle, we can repeat this procedure to determine $\varepsilon_0(\mathbf{r}_{\perp})$ for other collision configurations as long as we know the starting time of hydrodynamics $\tau_0$ and the energy density at that time $\varepsilon(\mathbf{r}_{\perp}=0,\tau_0)$. These two quantities are usually optimized to reproduce bulk experimental data after running hydrodynamic simulations. In the following analysis, however, we take a different approach in determining $\varepsilon_0(\mathbf{r}_{\perp})$ for other collision configurations. We assume that Eq. \eqref{eq:initial_energy_quoted} is the formula which determines the initial energy density $\varepsilon_0(\mathbf{r}_{\perp})$. Rewritten more explicitly, the initial energy density
\begin{equation}
\varepsilon_0(\mathbf{r}_{\perp})\equiv \varepsilon_0\left[\mu_i(\mathbf{r}_{\perp}, \sqrt{s}), Q_i(\mathbf{r}_{\perp},\sqrt{s})\right]
\end{equation}
is a functional of $\mu_i(\mathbf{r}_{\perp},\sqrt{s})$ and $Q_i(\mathbf{r}_{\perp},\sqrt{s})$ with $i=1,2$ indicating the two colliding nuclei. We assume the infrared cutoffs to be the same $m_1=m_2=\Lambda_{\rm QCD}$ for the two colliding nuclei. Both the color charge squared per unit area $\mu_i$ and the ultraviolet cutoff $Q_i$ depend, in principle, on the transverse position $\mathbf{r}_{\perp}$ and the center-of-momentum collision energy $\sqrt{s}$. The color charge squared per unit area for a nucleus is related to that of a nucleon by
\begin{equation}\label{eq:muA_TA_muN}
\mu_A(\mathbf{r},\sqrt{s}) = T_A(\mathbf{r}_{\perp})\mu_N(\sqrt{s}). 
\end{equation}
We assume, for a given nucleus-nucleus collision, that the ultraviolet cutoff is independent of the transverse position $\mathbf{r}_{\perp}$ so that
\begin{equation}\label{eq:Q_s_dependence}
Q_A(\mathbf{r}_{\perp},\sqrt{s}) =Q_A(\sqrt{s}). 
\end{equation}
The central Au+Au collision at $\sqrt{s_{NN}}= 200\,\rm{GeV}$ will serve as a reference for other collisions involving different nuclear sizes, different collision energies and different impact parameters. 

\subsection{Nuclear Size Dependence}

In this subsection, we consider heavy-ion collisions at fixed center-of-momentum collision energy $\sqrt{s_{NN}}=200\,\rm{GeV}$. We will study Cu+Cu, Cu+Au, and U+U collisions in addition to Au+Au collisions. RHIC has already run Cu+Cu, Au+Au, and Cu+Au collisions at $\sqrt{s_{NN}}=200\, \rm{GeV}$ and run U+U collisions at $\sqrt{s_{NN}}=193\, \rm{GeV}$.  For a fixed center-of-momentum collision energy, the initial energy density for a general nucleus-nucleus central collision can be obtained with reference to the Au+Au collision by
\begin{equation}\label{eq:initial_energy_equal_A}
\varepsilon_{0, A} (\mathbf{r}_{\perp}) = \left[\frac{T_A(\mathbf{r}_{\perp})}{T_{\rm Au}(\mathbf{r}_{\perp} =0)}\right]^2 \varepsilon_{0, \rm{Au}}(\mathbf{r}_{\perp}=0).
\end{equation}
The initial energy density for Au+Au collisions is $\varepsilon_{0, \rm{Au}}(\mathbf{r}_{\perp}=0) = 142.0\,\rm{GeV/fm}^3$ for the ultraviolet cutoff $Q=4.0\,\rm{GeV}$; see Fig. \ref{fig:enegy_vs_tau_diff_Q}.  Since the rapidity loss is insensitive to the ultraviolet cutoff, we choose $Q=4.0\,\rm{GeV}$ for all collisions at $\sqrt{s_{NN}} = 200\,\rm{GeV}$. 

Computation of nuclear thickness functions involves the Woods-Saxon nuclear distribution function. For spherical nuclei, such as Au and Cu, the Woods-Saxon distributions $\rho_A(r)$ are spherically symmetric. For deformed nuclei like U, we use the following parametrization \cite{Schenke:2014tga,Shen:2014vra}
\begin{equation}\label{eq:deformed_Woods_Saxon}
\rho_A(\mathbf{r}) = \frac{\rho_0}{1+e^{\frac{r-R_A\Omega(\theta)}{\xi}}}. 
\end{equation}
Here $\rho_0 = 0.166\,\rm{fm}^{-3}$, $\xi=0.44\,\rm{fm}$, $R_A=6.86\,\rm{fm}$ and $\Omega(\theta)=1+\beta_2Y_0^2(\theta)+\beta_4Y^4_0(\theta)$ with $\beta_2=0.280$ and $\beta_4=0.093$. The $Y^{l}_{m}(\theta)$ are spherical harmonic functions. The angle $\theta$ is related to the Cartesian coordinates by $\sin{\theta} = r_{\perp}/r$ and $\cos{\theta} = z/r$. The $\theta=0$ corresponds to the direction of the longest axis while $\theta=\pi/2$ corresponds to the direction of the shortest axis. Equation \eqref{eq:deformed_Woods_Saxon} describes an ellipsoid-like shape. In central U+U collisions, depending on the orientations of the uranium nuclei, there could be many different collision configurations. In the following discussion, we only consider the \textit{tip-tip} collision configuration where the longest axes of the uranium nuclei align with the beam directions. 

Table \ref{table:nuclear_size_dependence} presents several physical quantities associated with the three collisions of different nuclear sizes at $\sqrt{s_{NN}} = 200\,\rm{GeV}$, which are Cu+Cu, Au+Au and U+U (tip-tip) collisions. 
\begin{table}[bh]
\centering
{\renewcommand{\arraystretch}{1.3}
\begin{tabular}[t]{| c| c | c | c | } 
 \hline
$\sqrt{s_{NN}}=200\,\rm{GeV}$     & Cu+Cu  &  Au+Au  & U+U (tip-tip)\\
\hline\hline
$\varepsilon_{\rm hydro}(\rm{GeV/fm}^3)$ & 13.3 & 30.0 &  52.3 \\
 $\langle \delta y\rangle$    &  1.93 & 2.40  & 2.67\\
 $y_{\rm P}$    & 2.87   & 2.47      & 2.19 \\
$n_{\rm B}(\rm{1/fm}^3)$    & 2.04   & 3.01     & 3.94 \\ 
$\varepsilon_{\rm P}(\rm{GeV/fm}^3)$ & 9.0 & 20.0 & 33.8 \\ 
$T\,(\rm{MeV})$ & 264.1 & 328.1 & 376.0 \\ 
$\mu_{\rm B}(\rm{MeV})$ & 693.0 & 655.4 & 643.9 \\ 
$s/n_{\rm B}$ &18.33 & 23.53 & 27.82 \\
 \hline
\end{tabular}
}
\caption{Three different collision configurations at $\sqrt{s_{NN}} = 200\,\rm{GeV}$. Here $\varepsilon_{\rm hydro}$ is the initial energy density in the central rapidity region with $x=y=0$ at $\tau=0.6\,\rm{fm/c}$ when hydrodynamics starts. The average rapidity loss is $\langle \delta y\rangle$. The $y_{\rm P}$ denotes the final rapidity of the central core of the nucleus, which experiences the largest rapidity loss. Other thermodynamic quantities $n_{\rm B}, \varepsilon_{\rm P}, T$, $\mu_B$ and $s$ are also evaluated in the central core of the fireball. }
\label{table:nuclear_size_dependence}
\end{table}
They all have the same initial beam rapidity $y_0=5.36$. The average rapidity loss $\langle \delta y \rangle$ increases with the increase of nuclear mass. This can be understood from the tube-tube collision at the central cores of the colliding nuclei where the remaining quantities $y_{\rm P}$, $n_{\rm B}$, $\varepsilon_{\rm P}$, $T$ and $\mu_{\rm B}$ are evaluated. In the three collisions, the central core of a projectile starts with initial rapidity $y_0$ and ends with rapidity $y_{\rm P}$. For collisions involving nuclei of larger atomic mass, like U+U, the final rapidity $y_{\rm P}$ is smaller and the rapidity loss experienced is therefore larger. This point is encoded in the equations of motion \eqref{eq:slab_yM_eom} where $\mathcal{M}_{\rm P} \sim T_A(r_{\perp})$ and $\mathcal{A},\mathcal{B} \sim [T_A(r_{\perp})]^2$ so that the rate of rapidity change $dy/d\tau \sim T_A(r_{\perp})$. When averaging over $r_{\perp}$, collisions of nuclei with bigger nuclear size have larger average rapidity loss. The average rapidity loss in Au+Au collisions at $\sqrt{s_{NN}} = 200\,\rm{GeV}$ for the centrality class $0$-$5$\% has been measured and estimated to be in the range from 1.45 to 2.45 by the BRAHMS collaboration \cite{Bearden:2003hx}. However, the average rapidity losses for the Cu+Cu collision and the U+U (tip-tip) collision have not been measured experimentally. Baryon densities increase with increasing nuclear mass in accordance with the rapidity losses because the baryon density is proportional to the exponential of the rapidity loss. These maximal baryon densities are all more than ten times larger than the normal nuclear density. The maximal energy densities $\varepsilon_{\rm P}$ obtained in the fragmentation regions of the three collisions are smaller than the respective energy densities $\varepsilon_{\rm hydro}$ in the central rapidity region at $\tau=0.6\,\rm{fm/c}$ when hydrodynamics begins.  This is consistent with the expectation that energy density achieved in the central region of high energy heavy-ion collisions is larger than the energy density achieved in the receding nuclear fireballs. The temperature $T$ and baryon chemical potential $\mu_{\rm B}$ corresponding the largest baryon density $n_{\rm B}$ and energy density $\varepsilon_{\rm P}$ for the three collisions are also given in Table \ref{table:nuclear_size_dependence}. Finally, the largest entropy per baryon $s/n_B$ that can be achieved in the fragmentation regions of the three collisions are $18.33$, $23.53$ and $27.82$ which increases with the nuclear atomic mass, reflecting the fact that the energy density increases faster than the baryon density. 

\begin{figure}[thp]
 \centering
 \includegraphics[scale=0.93]{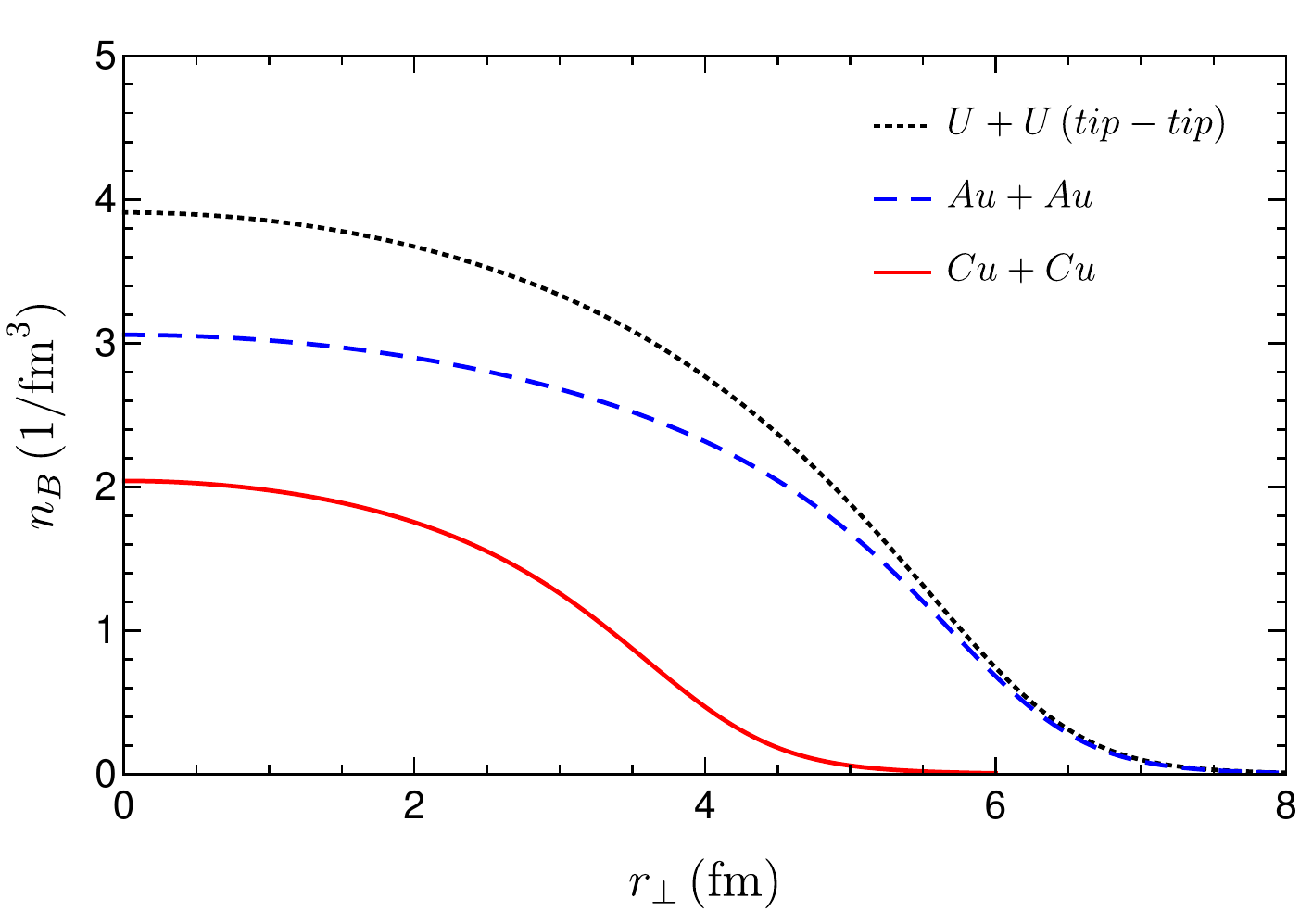}
 \caption{Baryon density achievable in the central core as a function of $r_{\perp}$ for three different collisions at $\sqrt{s_{NN}} = 200\,\rm{GeV}$. }
 \label{fig:compare_nB_rT_CuCu_AuAu_UU}
\end{figure}  

\begin{figure}[thp]
 \centering
 \includegraphics[scale=0.9]{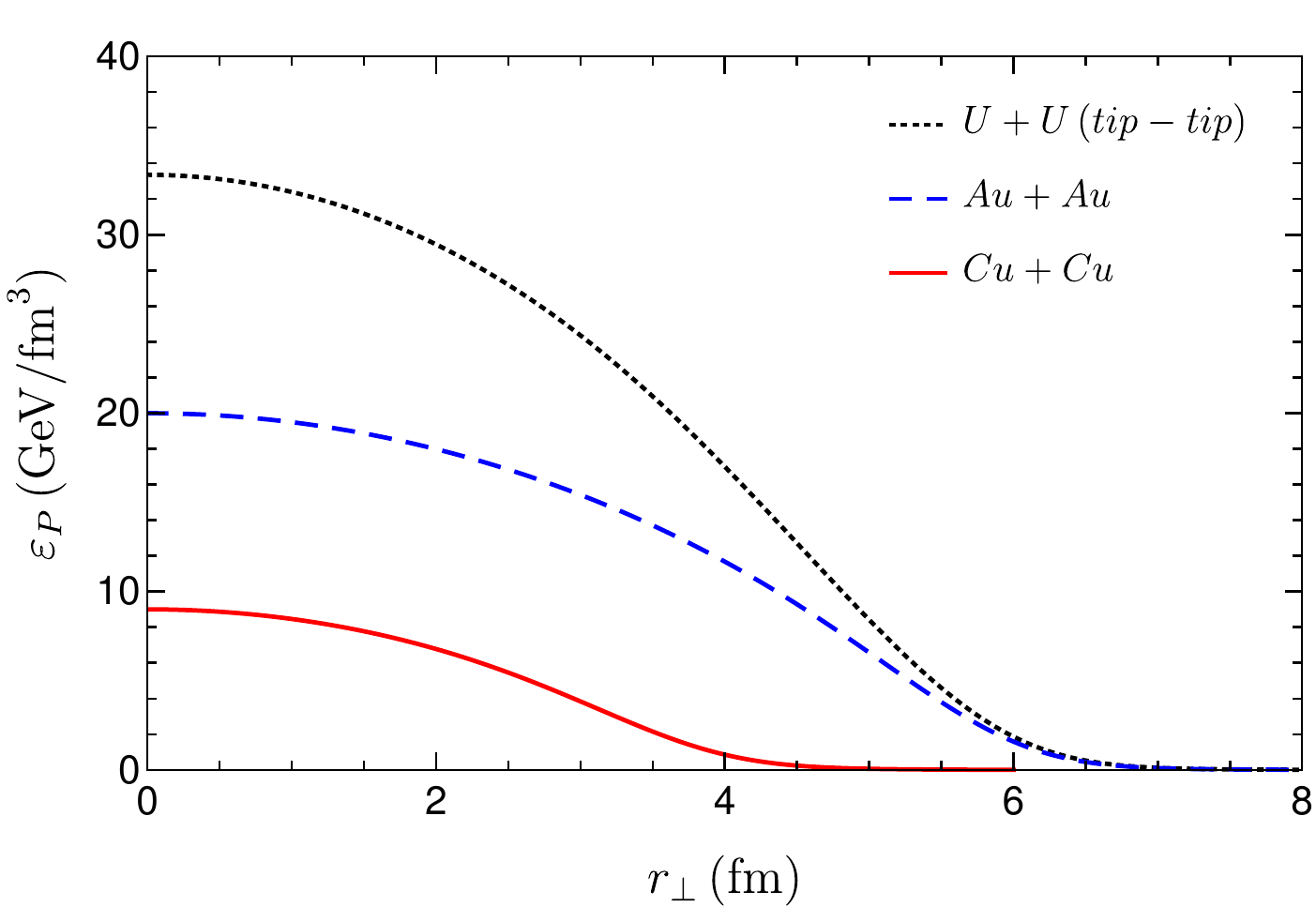}
 \caption{ Energy density achievable in the central core as a function of $r_{\perp}$ for three different collisions at $\sqrt{s_{NN}} = 200\,\rm{GeV}$. }
 \label{fig:compare_eP_rT_CuCu_AuAu_UU}
\end{figure}  

\begin{figure}[thp]
 \centering
 \includegraphics[scale=0.93]{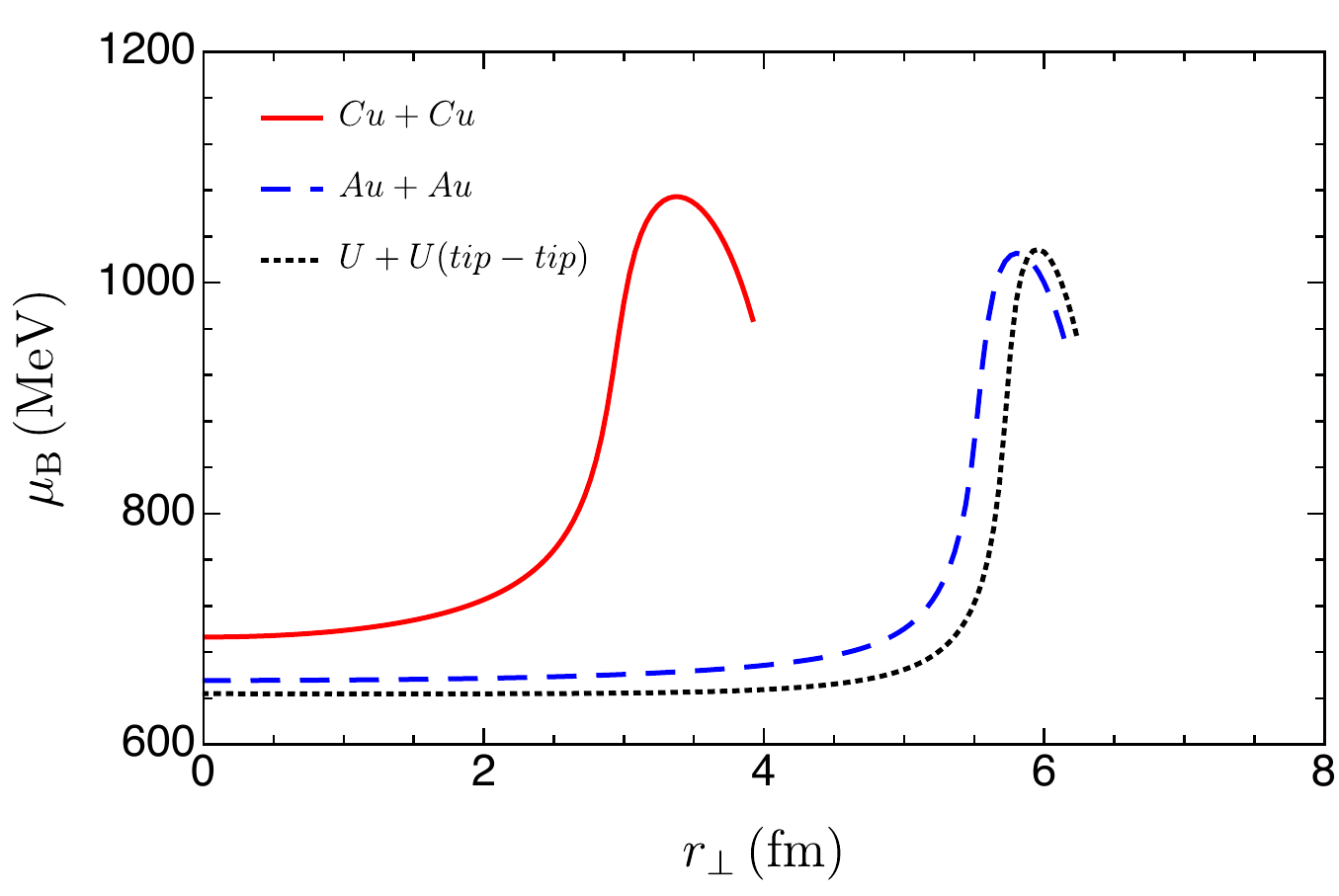}
 \caption{ Baryon chemical potential achievable in the central core as a function of $r_{\perp}$ for three different collisions at $\sqrt{s_{NN}} = 200\,\rm{GeV}$. Only regions with energy density larger than $1.0\,\rm{GeV/fm}^3$ have been displayed. }
 \label{fig:compare_MuB_rT_CuCu_AuAu_UU}
\end{figure}  

\begin{figure}[thp]
 \centering
 \includegraphics[scale=0.85]{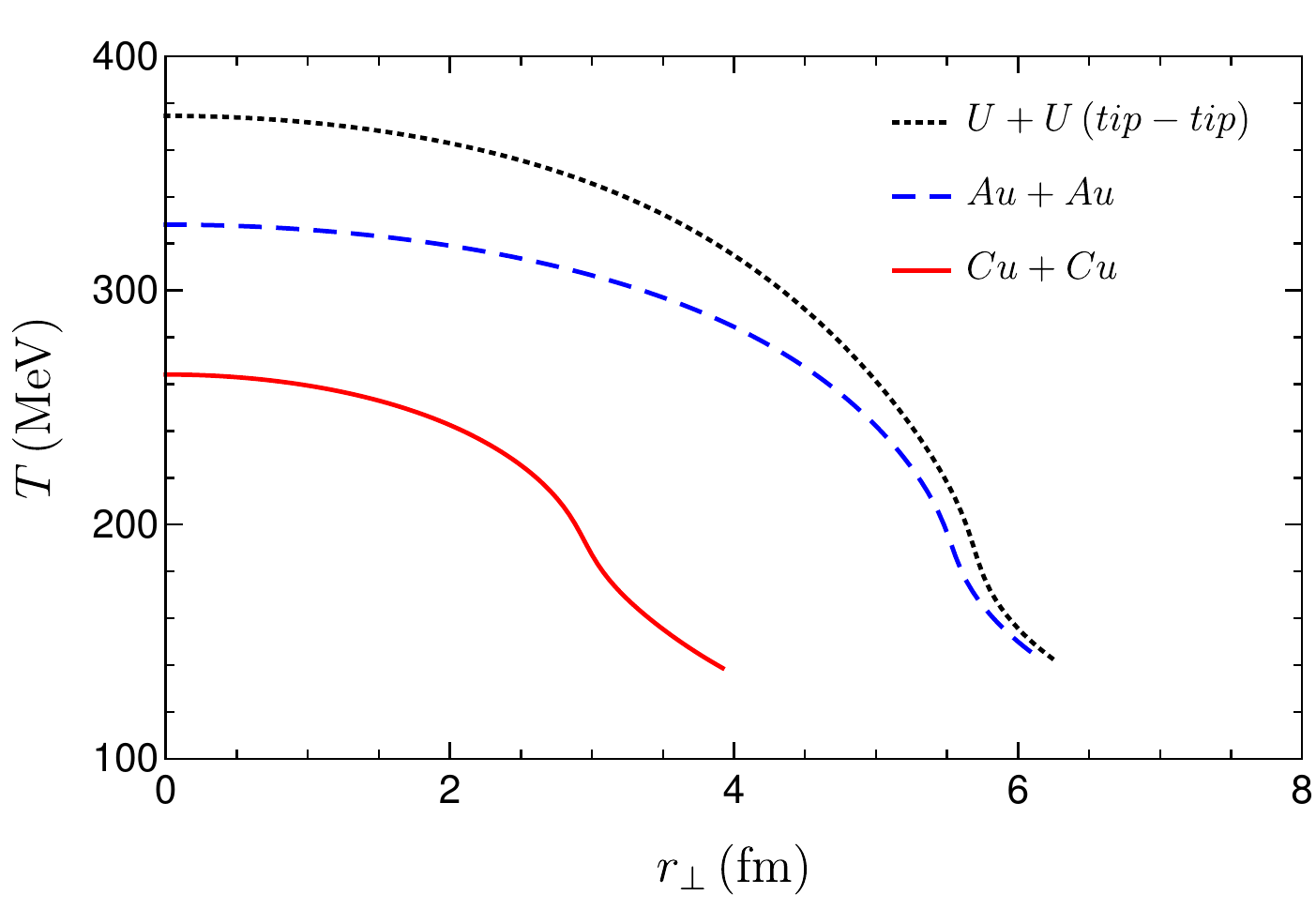}
 \caption{Temperature achievable in the central core as a function of $r_{\perp}$ for three different collisions at $\sqrt{s_{NN}} = 200\,\rm{GeV}$.  Only regions with energy density larger than $1.0\,\rm{GeV/fm}^3$ have been displayed.}
 \label{fig:compare_Temp_rT_CuCu_AuAu_UU}
\end{figure}  

\begin{figure}[thp]
 \centering
 \includegraphics[scale=0.9]{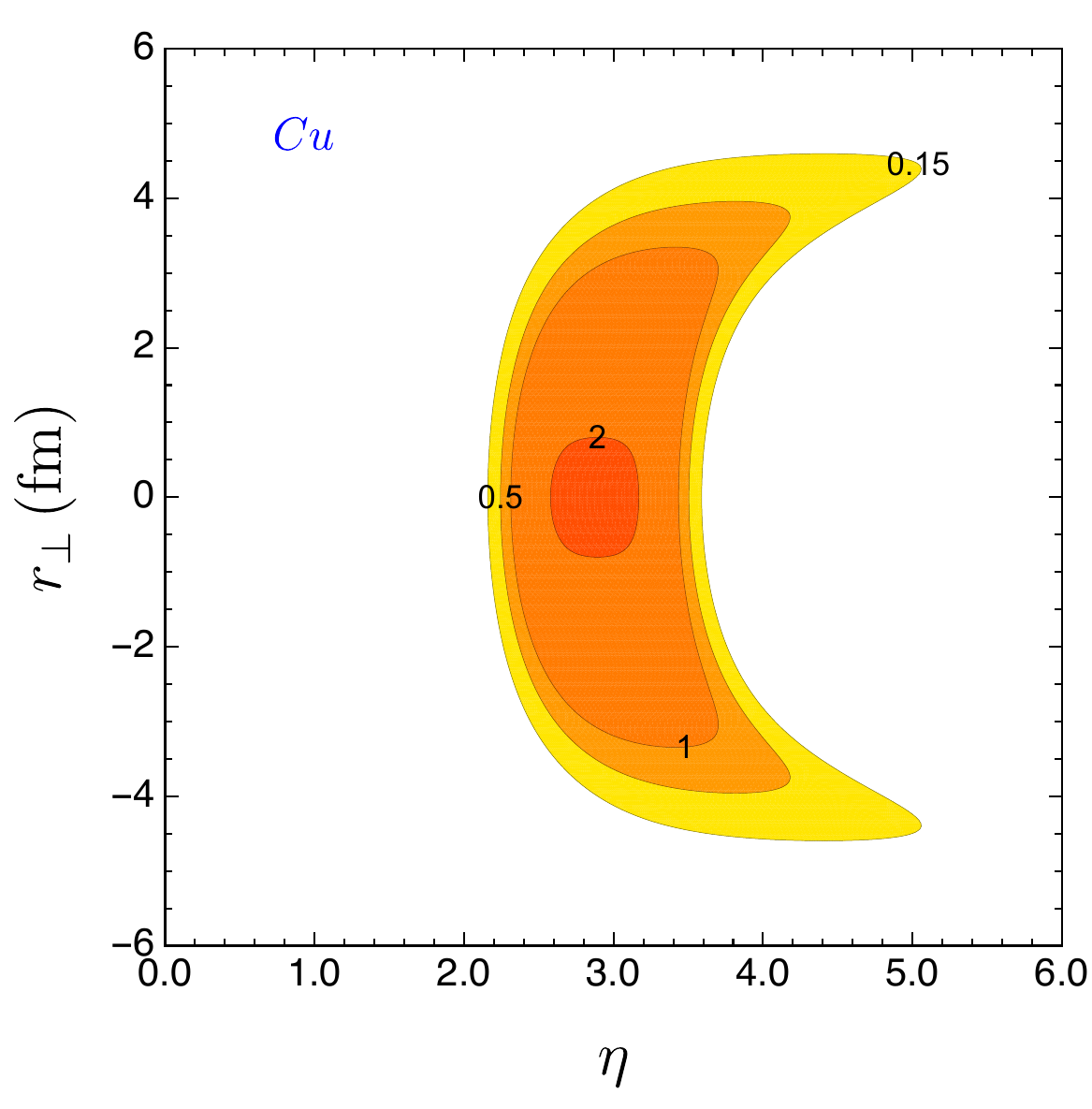}
 \caption{Contour plot of the proper baryon density for central collisions of Cu nuclei at $\sqrt{s_{NN}} = 200\, \rm{GeV}$. The numbers are in units of baryons per fm$^3$. The horizontal axis is the space-time pseudorapidity in the center-of-momentum frame.}
 \label{fig:rT_eta_nB_CuCu200}
\end{figure}

\begin{figure}[thp]
 \centering
 \includegraphics[scale=0.9]{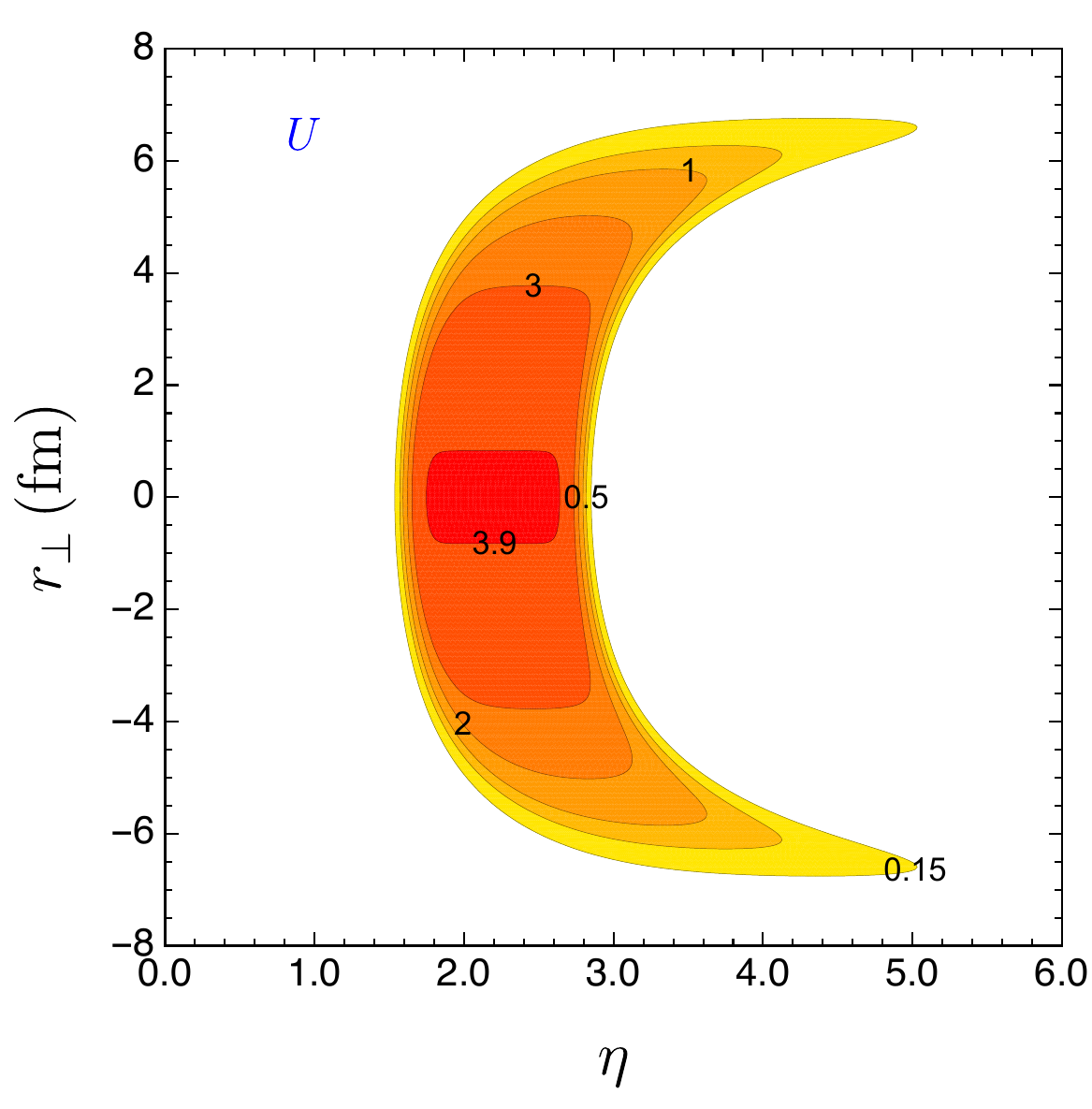}
 \caption{Contour plot of the proper baryon density for central tip-tip collisions of U nuclei at $\sqrt{s_{NN}} = 200\, \rm{GeV}$. The numbers are in units of baryons per fm$^3$. The horizontal axis is the space-time pseudorapidity in the center-of-momentum frame.}
 \label{fig:rT_eta_nB_UU200}
\end{figure}

Figures \ref{fig:compare_nB_rT_CuCu_AuAu_UU}, \ref{fig:compare_eP_rT_CuCu_AuAu_UU},  \ref{fig:compare_MuB_rT_CuCu_AuAu_UU} and \ref{fig:compare_Temp_rT_CuCu_AuAu_UU} show the baryon density, energy density, baryon chemical potential and temperature achievable in the central core of the three collisions as functions of $r_{\perp}$. The temperature and baryon chemical potential are displayed only in the regions with energy density larger than $1.0\,\rm{GeV/fm}^3$. In Fig. \ref{fig:compare_MuB_rT_CuCu_AuAu_UU}, the baryon chemical potential increases as $r_{\perp}$ increases, reaches a maximum, and then decreases. This feature is due to the equation of state because the corresponding temperatures in these regions are in the transition regions from the quark-gluon plasma to the hadronic resonance gas.   Figures \ref{fig:rT_eta_nB_CuCu200} and \ref{fig:rT_eta_nB_UU200} show the baryon density distributions in the center-of-momentum frame. Here $\eta$ is the space-time pseudorapidity. For Cu+Cu collisions, the pseudorapidity spans about $1.5$ unit around the central core regions while for U+U collision, the pseudorapidity spans about $1.0$ unit. 

For asymmetric central collisions of Cu+Au at $\sqrt{s_{NN}} = 200\,\rm{GeV}$, the gold nucleus has a larger diameter than the copper nucleus so that the transverse overlap size is only the cross sectional area of the copper nucleus.  Peripheral regions of the gold nucleus play the role of spectators and do not participate in the collision. To solve the equations of motion \eqref{eq:slab_yM_eom} for the copper and gold fireballs after the collision, one has to generalize the initial energy density in Eq. \eqref{eq:initial_energy_equal_A} to incorporate contributions from two different nuclei as
\begin{equation}
\varepsilon_{0, AA^{\prime}} (\mathbf{r}_{\perp}) = \frac{T_A(\mathbf{r}_{\perp})T_{A^{\prime}}(\mathbf{r}_{\perp})}{\left[T_{\rm Au}(\mathbf{r}_{\perp} =0)\right]^2}\, \varepsilon_{0, \rm{Au}}(\mathbf{r}_{\perp}=0).
\end{equation}
Here $A$ and $A^{\prime}$ indicate the two colliding nuclei. As a consequence, the functions $\mathcal{A},\mathcal{B}$ in the equations of motion \eqref{eq:slab_yM_eom} scale as  $T_A T_{A^{\prime}}$.  For the copper nucleus, after canceling $T_{\rm Cu}$ from the two sides of equations in Eq. \eqref{eq:slab_yM_eom}, the equations of motion only depend on the thickness function $T_{\rm Au}$. These equations of motion are exactly the same as the equations of motion for the gold nucleus in symmetric Au+Au collisions at the same energy. Likewise, the equations of motion for a gold nucleus in Cu+Au collisions are exactly the same as the equations of motion governing the copper nucleus in Cu+Cu collision at the same energy. In other words, concerning the rapidity loss and the excitation energy, the following equivalences are valid.
\vfill
\begin{enumerate}
\item[(i)] Copper in Cu+Au central collisions at $\sqrt{s_{NN}} = 200\,\rm{GeV}$  $\Longleftrightarrow$  Gold in Au+Au central collisions at $\sqrt{s_{NN}} = 200\,\rm{GeV}$.
\item[(ii)]  Gold in Cu+Au central collisions at $\sqrt{s_{NN}} = 200\,\rm{GeV}$  $\Longleftrightarrow$  Copper in Cu+Cu central collisions at $\sqrt{s_{NN}} = 200\,\rm{GeV}$.
\end{enumerate}
Therefore, the maximum baryon density achievable in the copper fireball in Cu+Au collisions is about $3.01\,\rm{baryons/fm}^3$ while the maximum baryon density in the gold fireball is about $2.04\,\rm{baryons/fm}^3$; compare Tables \ref{table:nuclear_size_dependence} and \ref{table:Cu_Au_collision}. The copper fireball is denser and hotter than the gold fireball in high energy Cu+Au collisions.  On the other hand, the average rapidity loss $\langle \delta y\rangle$ for the copper nucleus in Cu+Au collisions is different from that of the gold nucleus in Au+Au collisions because of the different nuclear thickness functions. 

\begin{table}[thp]
\centering
{\renewcommand{\arraystretch}{1.3}
\begin{tabular}[t]{| c| c | c | c | } 
 \hline
Cu+Au $\sqrt{s_{NN}}=200\,\rm{GeV}$  & Au & Cu \\
\hline\hline
$R_{A}(\rm{fm})$ & 6.4 & 4.2 \\
 $\langle \delta y\rangle$    &  1.34 & 2.71\\
 $y_{\rm P}$    & 2.87   & 2.47     \\
$n_{\rm B}(\rm{1/fm}^3)$    & 2.04   & 3.01    \\ 
$\varepsilon_{\rm P}(\rm{GeV/fm}^3)$ & 9.0 & 20.0  \\ 
$T\,(\rm{MeV})$ & 264.1 & 328.1  \\ 
$\mu_{\rm B}(\rm{MeV})$ & 693.0 & 655.4  \\ 
$s/n_{\rm B}$ &18.33 & 23.53 \\
 \hline
\end{tabular}
}
\caption{Asymmetric Cu+Au collisions at $\sqrt{s_{NN}} = 200\,\rm{GeV}$. 
The average rapidity loss is $\langle \delta y\rangle$. The $y_{\rm P}$ denotes the final rapidity of the central core of the nucleus, which experiences the largest rapidity loss. Other thermodynamic quantities $n_{\rm B}, \varepsilon_{\rm P}, T$, $\mu_B$ and $s$ are also evaluated in the central core of the fireball.   }
\label{table:Cu_Au_collision}
\end{table}

\begin{figure}[thp]
 \centering
 \includegraphics[scale=1.0]{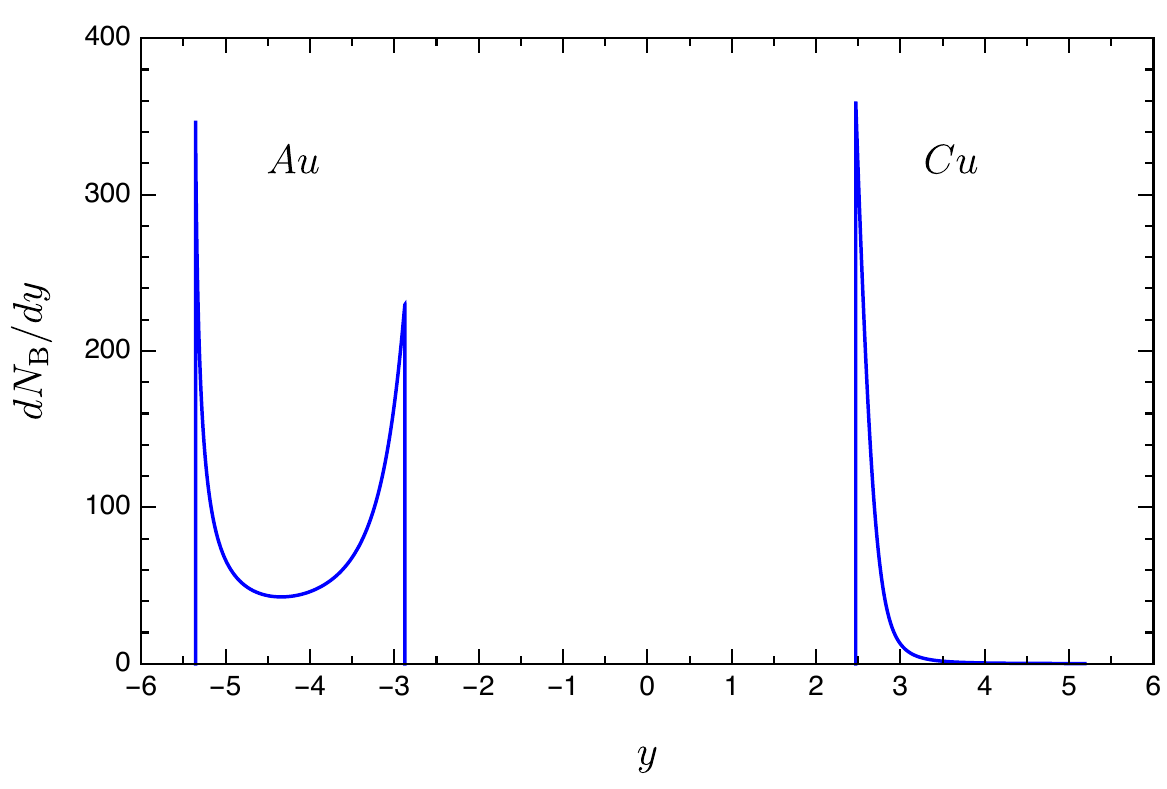}
 \caption{Net baryon rapidity distribution in Cu+Au central collisions at $\sqrt{s_{NN}} = 200\,\rm{GeV}$. The initial beam rapidities are $\pm 5.36$. Peripheral regions of the Au nucleus do not participate in the collision.}
 \label{fig:dNBdy_CuAu200}
\end{figure}  

Figure \ref{fig:dNBdy_CuAu200} shows the net baryon rapidity distribution in Cu+Au central collisions at $\sqrt{s_{NN}} = 200\,\rm{GeV}$; this is the initial distribution to be used in hydrodynamic modeling. Peripheral regions of the gold nucleus do not participate in the collision so that their rapidities are the same as the beam rapidity $y_0=-5.36$. 
The net baryon distribution on the copper side is smaller magnitude and narrower when compared to the net baryon distribution in Au+Au collisions as shown in Fig. \ref{fig:dNB_dy_vs_y}, but they have the same smallest rapidity cutoff at $y=2.47$. 

\subsection{Collision Energy Dependence}

In this section we study the high baryon densities achievable with different collision energies. Specifically, we focus on Au+Au central collisions at $\sqrt{s_{NN}} = 62.4\,\rm{GeV}$, Au+Au central collisions at $\sqrt{s_{NN}} = 200\,\rm{GeV}$, Pb+Pb central collisions at $\sqrt{s_{NN}} = 2.76\,\rm{TeV}$ and Pb+Pb central collisions at $\sqrt{s_{NN}} = 5.02\,\rm{TeV}$. The Au+Au collisions at these two different collision energies have been carried out at RHIC while the LHC has collided Pb+Pb at $2.76\,\rm{TeV}$ and $5.02\,\rm{TeV}$. The Au nucleus has atomic mass number 197 and the Pb nucleus has atomic mass number 208, so nuclear size differences would be minor. It should be pointed out that the McLerran-Venugopalan model that describes the central region of high energy heavy-ion collisions using classical gluon fields becomes more applicable with increasing center-of-momentum collision energy. For the Au+Au collision at $\sqrt{s_{NN}} = 62.4\,\rm{GeV}$, whether the McLerran-Venugopalan model is still applicable needs more detailed study, which is beyond the scope of this paper. Here we explore the Au+Au collision at $\sqrt{s_{NN}}=62.4\,\rm{GeV}$ as an extrapolation of our method. Importantly, the average baryon rapidity loss and net proton rapidity distribution have been measured for Au+Au collision at $\sqrt{s_{NN}} = 62.4\,\rm{GeV}$ \cite{Arsene:2009aa}. 

To solve for the rapidity loss and excitation energy from Eq. \eqref{eq:slab_yM_eom} for collisions at different energies, one needs to generalize the calculation of initial energy density from Eq. \eqref{eq:initial_energy_equal_A} to include the collision energy dependence. With the help of Eq. \eqref{eq:muA_TA_muN} we have
\begin{equation}
\frac{\varepsilon_{0, A}(r_{\perp}, \sqrt{s_{NN}})}{\varepsilon_{0,Au}(r_{\perp}=0,\sqrt{s_{NN}}=200\,\rm{GeV})} = \left[\frac{T_A(r_{\perp})}{T_{Au}(r_{\perp}=0)}\right]^2\left[\frac{\mu_{N}(\sqrt{s_{NN}})}{\mu_N(\sqrt{s_{NN}}=200\,\rm{GeV})}\right]^2 \,.
\end{equation}
In obtaining this expression, we ignore the expected collision energy dependence of the ultraviolet cutoff illustrated in Eq. \eqref{eq:Q_s_dependence} which appears as $\ln{Q^2/m^2}$ in the initial energy density expression of Eq. \eqref{eq:initial_energy_quoted}. In principle the ultraviolet cutoff $Q$ has to be adjusted when the saturation scale $Q_s$ changes with collision energy. This is to make sure that the scale separation $Q^2\gg Q_s^2$ is satisfied so that the semi-analytic expression of the glasma energy-momentum tensor we obtained in the leading $Q^2$ approximation can be used.  In Au+Au collisions at $\sqrt{s_{NN}}=200\,\rm{GeV}$, we used the value $Q_s=1.2\,\rm{GeV}$ and $Q=4.0\,\rm{GeV}$. As it will become clear in the following, the saturation scale at the LHC energy $5.02\,\rm{TeV}$ is approximately $Q_s \sim 1.8\,\rm{GeV}$ and an ultraviolet cutoff of $Q\sim 6.0\,\rm{GeV}$ only contributes approximately $10\%$ change in the energy density after taking the logarithm. This logarithmic change of the ultraviolet cutoff in the expression for the energy density should be minor compared to the power law changes in the $\mu_N(\sqrt{s_{NN}})$. 
Note that $\mu_N$ is related to the saturation scale of the nucleon $Q_{sN}^2$ up to a logarithmic correction. The collision energy dependence of the saturation scale $Q_{sN}^2(x)$ can be parameterized as \cite{GolecBiernat:1998js,GolecBiernat:1999qd,Stasto:2000er}
\begin{equation}
Q_{sN}^2(x) = Q_0^2\left(\frac{x_0}{x}\right)^{\lambda}
\end{equation}
with $Q_0^2 = 1.0\,\rm{GeV}^2$, $x_0=3\times 10^{-4}$ and $\lambda = 0.288$. This parametrization provides a successful description of the HERA data on deep inelastic scattering. At central rapidity $y=0$, the small $x$ parameter is related to the center-of-momentum collision energy by $x=Q_{sN}/\sqrt{s_{NN}}$, and  one obtains \cite{Kharzeev:2004if}
\begin{equation}
\frac{Q_{sN}^2(\sqrt{s_{NN}})}{Q_{sN}^2(\sqrt{s_0})} = \left(\frac{\sqrt{s_{NN}}}{\sqrt{s_0}}\right)^{\bar{\lambda}}
\end{equation}
where $\bar{\lambda} = \frac{\lambda}{1+\lambda/2}=0.252$ with $\sqrt{s_0}$ some reference collision energy. Therefore, the final expression for the collision energy dependent initial energy density is 
\begin{equation}\label{eq:collision_energy_dependence}
\frac{\varepsilon_{0, A}(r_{\perp}, \sqrt{s_{NN}})}{\varepsilon_{0,Au}(r_{\perp}=0,\sqrt{s_{NN}}=200\,\rm{GeV})} = \left[\frac{T_A(r_{\perp})}{T_{Au}(r_{\perp}=0)}\right]^2\left[\frac{\sqrt{s_{NN}}}{200\,\rm{GeV}}\right]^{2\bar{\lambda}}.
\end{equation}
Using this simple parametrization of the collision energy dependence, we compute the initial energy density of the glasma and the hydrodynamic initial energy density for the different collision energies; the results are given in Table \ref{table:collision_energy_dependence_one}.

\begin{table}[thp]
\centering
{\renewcommand{\arraystretch}{1.3}
\begin{tabular}[t]{| c| c | c | c | c | } 
 \hline
Collision    & Au+Au & Au+Au & Pb+Pb & Pb+Pb  \\
\hline\hline
$\sqrt{s_{NN}} (\rm{GeV})$  & 62.4  & 200 & 2760 & 5020     \\ 
$Q_{sA}$(\rm{GeV}) & 1.04 & 1.20 & 1.69 & 1.82 \\
$\varepsilon_0 (\rm{GeV/fm}^3)$                  & 78.9 & 142.0 & 553.6 & 748.5\\
$\varepsilon_{\rm hydro}(\rm{GeV/fm}^3)$          & 16.7 & 30.0 & 116.9 & 158.0\\
$\varepsilon^{\rm MC-Glauber}_{\rm hydro}(\rm{GeV/fm}^3) $ & 25.5 & 42.5 & 104.5 & 132.3\\
 \hline
\end{tabular}
}
\caption{The initial energy density of the glasma and the initial energy density for hydrodynamics for different collision energies from Eq. \eqref{eq:collision_energy_dependence}. The hydrodynamic initial energy densities obtained from the Monte Carlo Glauber model for the 0-5\% centrality are from Ref. \cite{ChunShen2016}. The gluon saturation scale of the nucleus at different collision energies are also estimated. }
\label{table:collision_energy_dependence_one}
\end{table}

\begin{figure}[bhp]
 \centering
 \includegraphics[scale=0.76]{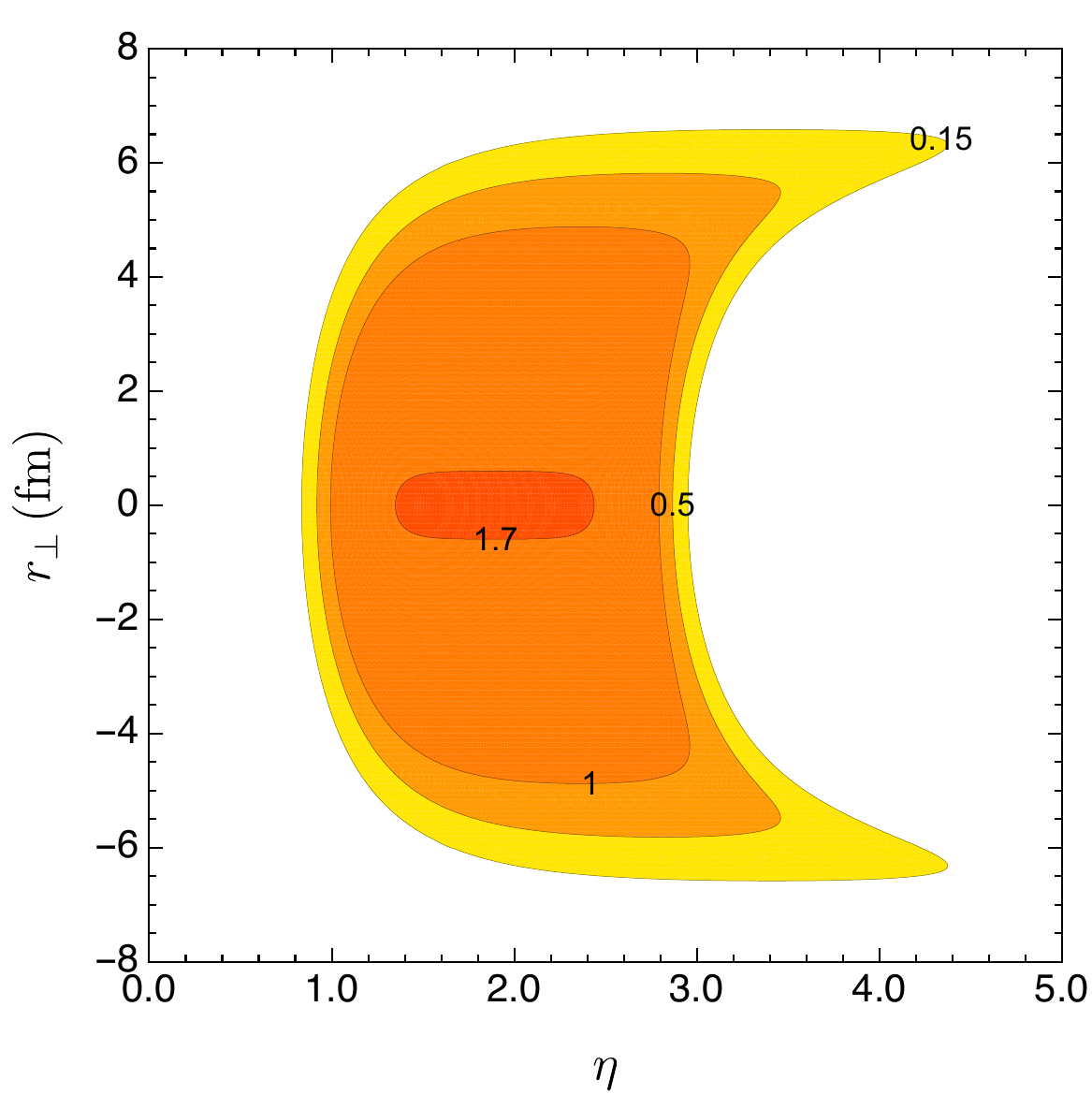}
 \caption{Contour plot of the proper baryon density for central collisions of gold nuclei at $\sqrt{s_{NN}} = 62.4\, \rm{GeV}$. The numbers are in units of baryons per fm$^3$. The horizontal axis is the space-time pseudorapidity in the center-of-momentum frame.}
 \label{fig:nB_rT_z_AuAu62p4}
\end{figure}

\begin{figure}[thp]
 \centering
 \includegraphics[scale=0.72]{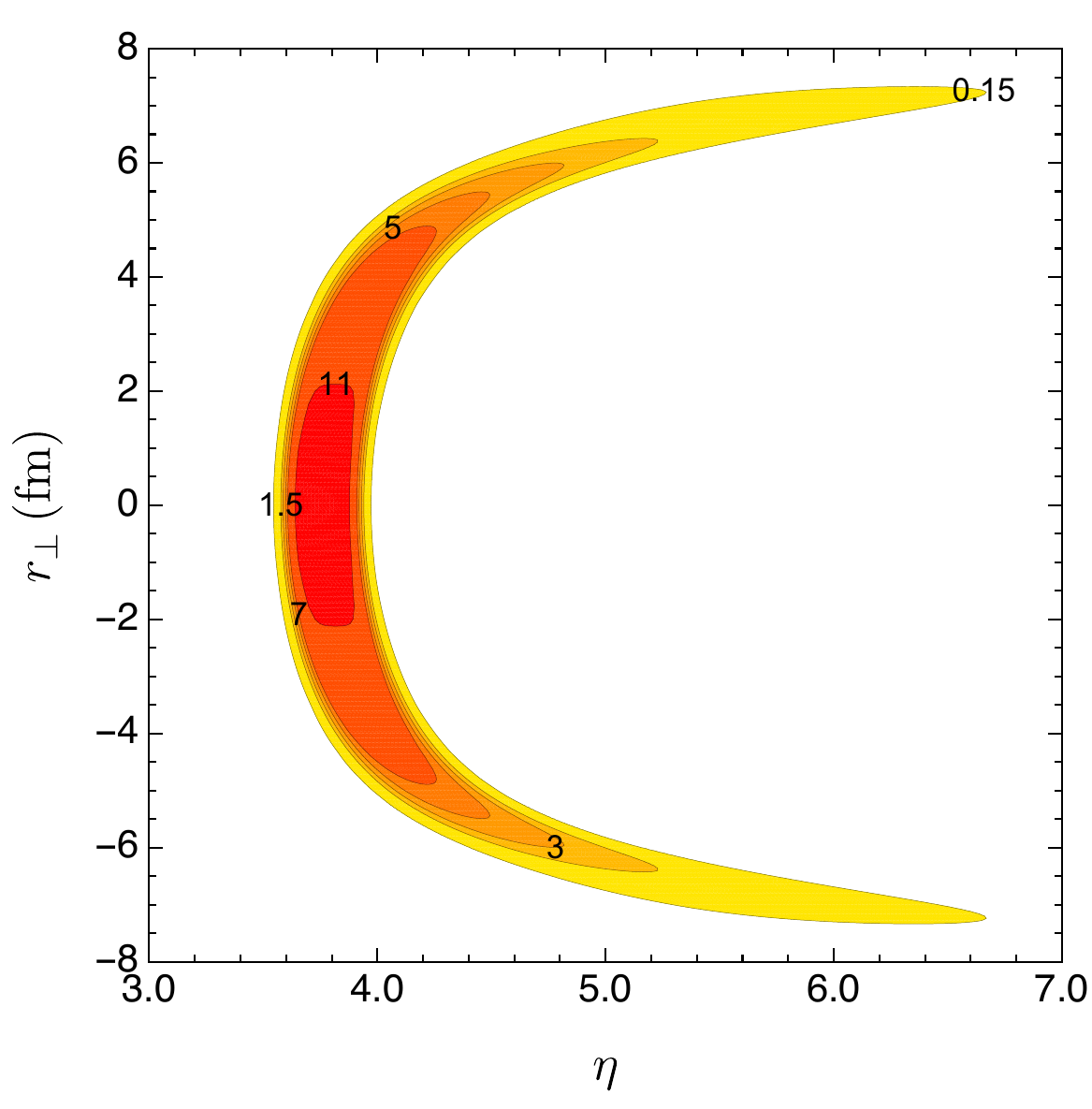}
 \caption{Contour plot of the proper baryon density for central collisions of lead nuclei at $\sqrt{s_{NN}} = 2.76\, \rm{TeV}$. The numbers are in units of baryons per fm$^3$. The 
horizontal axis is the space-time pseudorapidity in the center-of-momentum frame.}
 \label{fig:nB_rT_z_PbPb@2760GeV}
\end{figure}

With the help of Eq. \eqref{eq:collision_energy_dependence} we can calculate the baryon and energy densities achieved in Au+Au and Pb+Pb collisions at different energies. Table \ref{table:collision_energy_dependence_two} shows a few characteristic values of physical quantities. 
\begin{table}[thp]
\centering
{\renewcommand{\arraystretch}{1.3}
\begin{tabular}[t]{| c| c | c | c | c | } 
 \hline
Collision    & Au+Au & Au+Au & Pb+Pb & Pb+Pb  \\
\hline\hline
$\sqrt{s_{NN}} (\rm{GeV})$  & 62.4  & 200 & 2760 & 5020     \\ 
$y_0$                                    & 4.2 & 5.36 & 7.99 & 8.59 \\
$\langle \delta y\rangle$     & 1.85 & 2.41 & 3.73 & 4.02\\
$y_{\rm P}$                          & 1.89 & 2.47 & 3.76 & 4.06\\
$n_{\rm B}(\rm{1/fm}^3) $ & 1.71 & 3.01 & 11.66 & 15.77\\
$\varepsilon_{\rm P}(\rm{GeV/fm}^3) $ & 6.34 & 20.00 & 288.67 & 527.50 \\
$T(\rm{MeV})$  & 236.6 & 328.1 & 642.4 & 745.9 \\
$\mu_{\rm B}(\rm{MeV}) $ &748.3 & 655.4 & 673.7 & 680.8 \\
$s/n$ & 16.3 & 23.5 & 49.6 & 58.2 \\ 
 \hline
\end{tabular}
}
\caption{Central collisions of Au+Au at $\sqrt{s_{NN}}$ = 62.4 and 200 GeV and of Pb+Pb at $\sqrt{s_{NN}}$ = 2.76 and 5.02TeV. The final rapidity $y_{\rm P}$, baryon density $n_{\rm B}$ , energy density $\varepsilon_{\rm P}$ , temperature $T$, baryon chemical potential $\mu_B$ and the entropy per baryon ratio $s/n$ are given for the central core of the fireball. }
\label{table:collision_energy_dependence_two}
\end{table}
The average rapidity loss for Au+Au at $\sqrt{s_{NN}} = 62.4 \,\rm{GeV}$ is found to be 1.85, which is close to the lower bound of $1.85\leq \langle \delta y \rangle \leq 2.17$ measured by the BRAHMS collaboration \cite{Arsene:2009aa}. The average rapidity losses for Pb+Pb collisions at $2.76\,\rm{TeV}$ and $5.02\,\rm{TeV}$ are 3.73 and 4.02, respectively. As a consequence of the rapidity losses, the maximum baryon density achievable in Au+Au collisions at $\sqrt{s_{NN}} = 62.4\,\rm{GeV}$ is 1.7 baryons/fm$^3$, which is about 11 times larger than normal nuclear density. Furthermore, for Pb+Pb collisions at $\sqrt{s_{NN}} = 2.76\,\rm{TeV}$ and $5.02\,\rm{TeV}$, the maximum baryon densities achievable are 11.7 baryons/fm$^3$ and 15.8 baryons/fm$^3$, which are about 75 times and 101 times larger than normal nuclear density. These are extremely large baryon densities. The maximum energy density in the fireball for Au+Au collisions at $\sqrt{s_{NN}} = 62.4\,\rm{GeV}$ is $6.34\,\rm{GeV/fm}^3$ which is smaller than the energy density in the central region $16.7\,\rm{GeV/fm}^3$, as shown in Table \ref{table:collision_energy_dependence_one}. However, the energy density in the receding fireballs in Pb+Pb collisions at $\sqrt{s_{NN}} = 2.76\,\rm{TeV}$ and $5.02\,\rm{TeV}$ are $288.7\,\rm{GeV/fm}^3$ and $527.5\,\rm{GeV/fm}^3$ which are much larger than the corresponding energy densities in the central region, $117\,\rm{GeV/fm}^3$ and $ 158\,\rm{GeV/fm}^3$ given in Table \ref{table:collision_energy_dependence_one}.  As the collision energy increases, the maximum temperature achievable in the nuclear fireballs increases monotonically. The baryon chemical potential, however, increases rather slowly.  Consequently, the maximum entropy per baryon increases with the collision energy, as one would expect.  This is a direct result of the use of the McLerran-Venugopalan model for the glasma phase.
The baryon distributions of the fireballs for Au+Au collisions at $\sqrt{s_{NN}} = 62.4\,\rm{GeV}$ and for Pb+Pb collisions at $\sqrt{s_{NN}} = 2.76\,\rm{TeV}$ are shown in Figs. \ref{fig:nB_rT_z_AuAu62p4} and \ref{fig:nB_rT_z_PbPb@2760GeV}. 

\begin{figure}[bhp]
 \centering
 \includegraphics[scale=0.79]{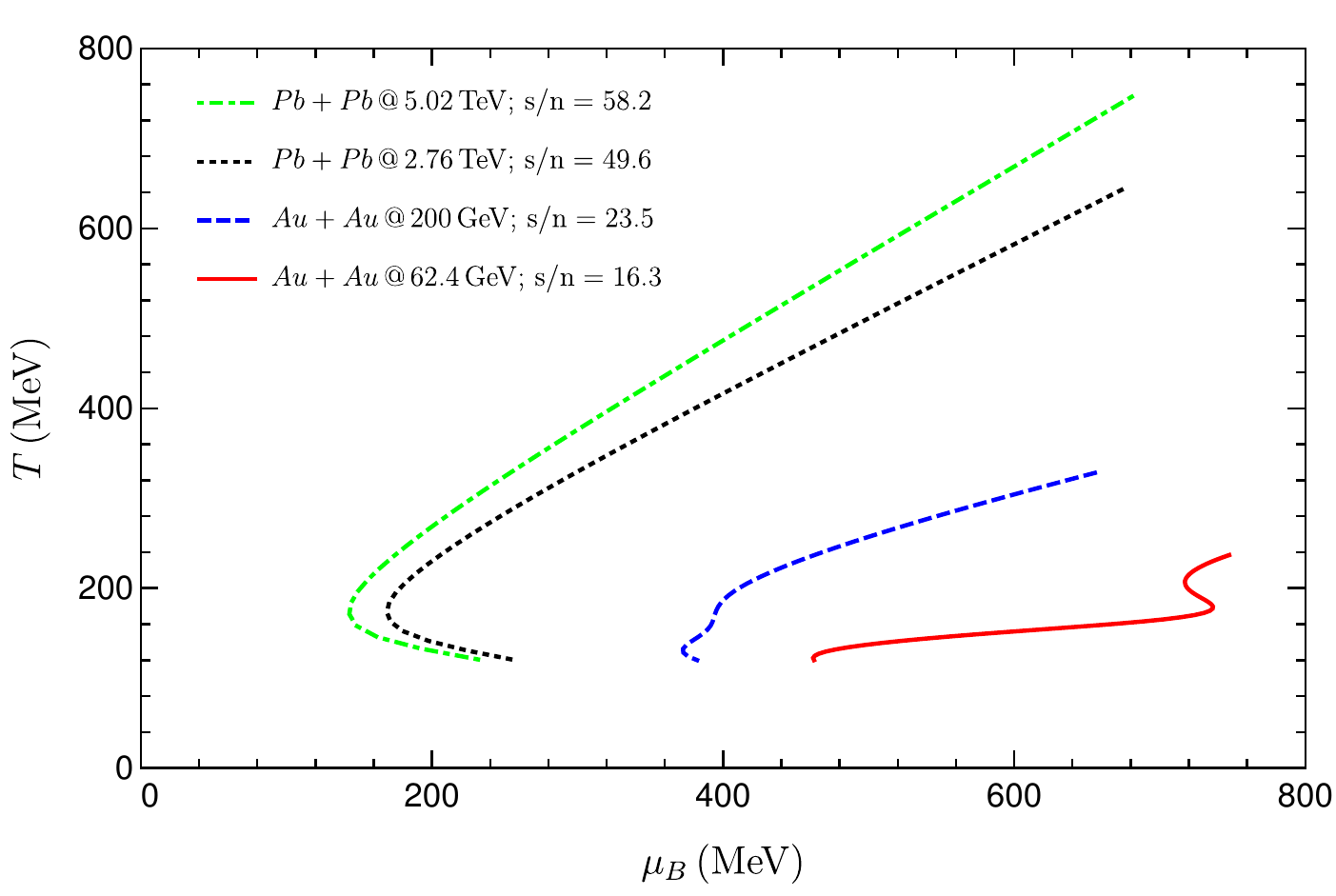}
 \caption{Adiabatic phase trajectories for the central cores of the fireball in Au+Au collisions at $\sqrt{s_{NN}}$ = 62.4 and 200 GeV and in Pb+Pb collisions at $\sqrt{s_{NN}}$ = 2.76 and 5.02 TeV.}
 \label{fig:phase_trajectory_diff_s}
\end{figure}

In Fig. \ref{fig:phase_trajectory_diff_s} the adiabatic phase trajectories for the central cores of the fireballs for Au+Au collisions with $\sqrt{s_{NN}}$ = 62.4 and 200 GeV and for Pb+Pb collisions with $\sqrt{s_{NN}}$ = 2.76 and 5.02 TeV are shown. Increasing the collision energy increases the temperatures while the baryon chemical potentials change more slowly, leading to larger entropy per baryon ratios.  

\subsection{Non-Central Collisions}

Apart from colliding heavy ions of different sizes and collision energies, realistic heavy-ion collisions also measure observables at different centralities. The centrality characterizes the impact parameter of the two colliding nuclei. From the most central collisions with impact parameter $b \sim 0$ to the most peripheral collisions with impact parameter $b \sim 2R_A$, the centrality increases from $0\%$ to $100\%$. The impact parameter, however, cannot be directly measured in heavy-ion experiments.  It is to be noted that the observed particle multiplicity decreases from central collisions to peripheral collisions. Therefore, experiments measure the particle multiplicity and categorize these events using centrality values in the range 0\%-100\$. With the help of a geometric model, like the Glauber model, one can then relate the centrality to the impact parameter. A few examples are given in  Table \ref{table:centrality_impact_parameter} as calculated by the PHENIX collaboration \cite{Reygers}.  In this subsection, we will explore the impact parameter dependence of the baryon densities in the fireballs in high energy heavy-ion collisions. 

\begin{table}[thp]
\centering
{\renewcommand{\arraystretch}{1.3}
\begin{tabular}[t]{| c| c | c | c | c | c | c| c|} 
 \hline
Centrality Class    & 0-5\% & 5\%-10\% & 10\%-15\% & 15\%-20\% & 20\%-25\%  \\
\hline
Impact parameter (fm)  & 2.3  & 4.1 & 5.2 & 6.2 & 7.0     \\ 
 \hline
\end{tabular}
}
\caption{Correspondence between centrality classes and average impact parameters from calculations within the Glauber model.}
\label{table:centrality_impact_parameter}
\end{table}

\begin{figure}[thp]
 \centering
 \includegraphics[scale=0.8]{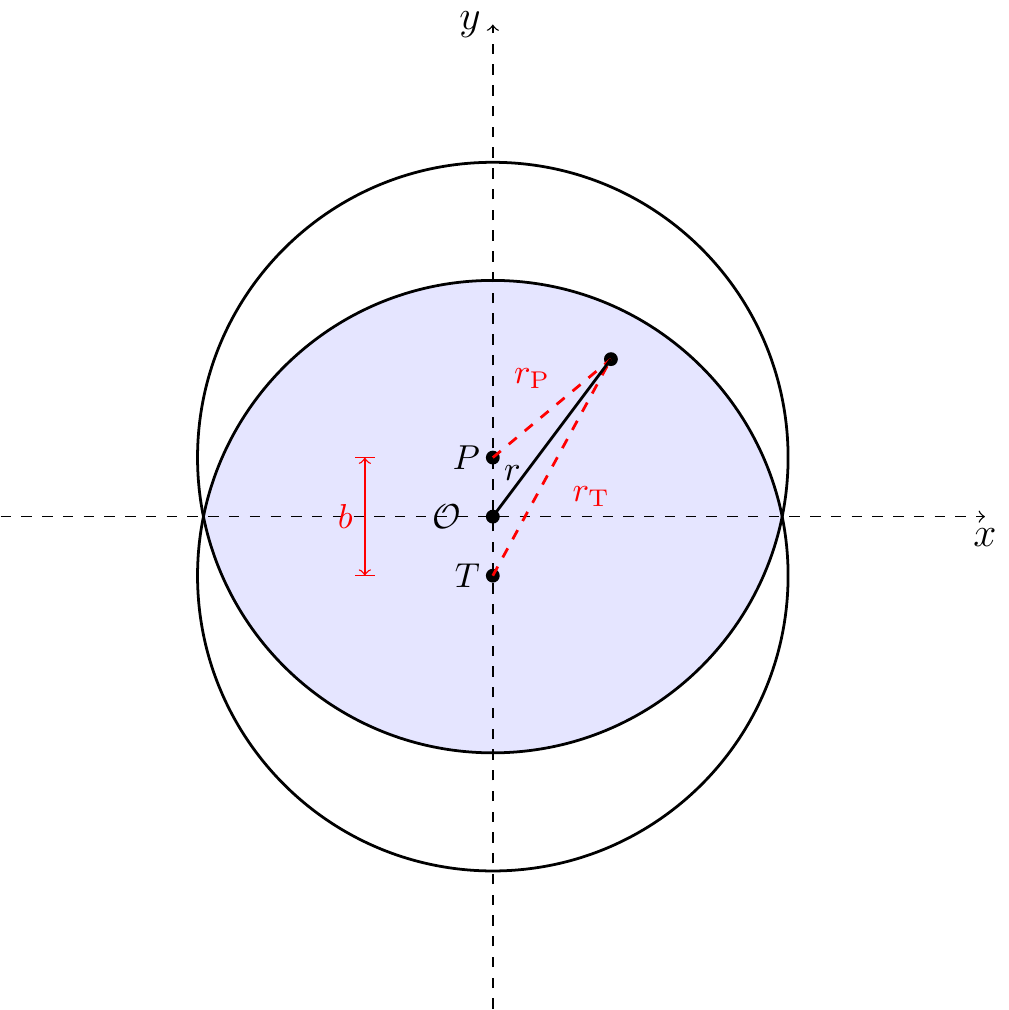}
 \caption{Schematic picture of the transverse overlap region for non-central collisions of two nuclei.}
 \label{fig:non_central_illustration}
\end{figure}

To characterize non-central collisions in the transverse plane we use cylindrical coordinates.   Figure \ref{fig:non_central_illustration} is a schematic illustration of the transverse overlap region of two equal size nuclei colliding at non-zero impact parameter $b$. All the vectors in this figure are two-dimensional in the $x$-$y$ plane. Let the projectile P be located at $\mathbf{b}/2$ and the target T at $-\mathbf{b}/2$ with $\mathbf{b} = (0,b)$. An arbitrary point is labeled by $\mathbf{r}_{\perp} = r_{\perp}(\cos{\phi}, \sin{\phi}) = (x,y)$.  The distance from the centers of the nuclei to that point are
\begin{equation}
\begin{split}
&r_{\rm P}^2 = r_{\perp}^2 + \frac{1}{4}b^2 - br_{\perp}\sin{\phi} \, ,\\
&r_{\rm T}^2 = r_{\perp}^2 + \frac{1}{4}b^2 +br_{\perp}\sin{\phi} \, .
\end{split}
\end{equation}
The thickness functions $T_{\rm P}$ and $T_{\rm T}$ depend only on $r_{\rm P}$ and $r_{\rm T}$, respectively. Consider the baryon distribution arising from the projectile; a similar formula applies to the target. Let $y_{\rm P}(r_{\perp},\phi, b)$ denote the final rapidity of a piece of projectile located at the position $\mathbf{r}_{\perp}$. Then the baryon rapidity distribution is a generalization of Eq. \eqref{eq:netB_dis_no_thermal} taking into account the non-zero impact parameter,
\begin{equation}
\frac{dN_P}{dy} = \int_0^{\infty}dr\int_0^{2\pi} d\phi T_{\rm P}(r_{\perp},\phi,b) r_{\perp}\delta(y-y_{\rm P}(r_{\perp},\phi, b)). 
\end{equation}
For central collisions of spherical nuclei, which need not be identical, there is no dependence on $\phi$, and there is a one-to-one correspondence between $y_{\rm P}$ and $r$. Then we can replace $\delta(y-y_{\rm P}(r_{\perp}))$ with $\delta(r_{\perp} - r_{\perp\rm P}(y))$ along with the relevant Jacobian to reproduce the expression in Eq.  \eqref{eq:netB_dis_no_thermal}. 

For $b\neq 0$ it is may be better to discretize both $r$ and $\phi$. Assume uniform spacings $\Delta r$ and $\Delta \phi$ which satisfy $N_r \times r = \tilde{R}$ and $(N_{\phi}+1)\times \Delta \phi = 2\pi$, respectively. Here $\tilde{R}$ characterizes the radius of a circular area chosen for the discretization. The discretized points are labeled by $(r_i, \phi_j)$ with
\begin{equation}
\begin{split}
& r_i = i\Delta r, \qquad i = 1,2, \ldots, N_r\\
&\phi_j = j\Delta \phi, \qquad j=0,1,2, \ldots, N_{\phi}. 
\end{split}
\end{equation}
The point $(0,0)$ is treated separately. For points in the overlap region, their coordinates $(r_i, \phi_j)$ satisfy $r^2_{\mathrm{P}(i,j)}\leq R_A^2$ and $r^2_{\mathrm{T}(i,j)}\leq R_A^2$. For the tube-tube collisions at the position $(r_i, \phi_j)$, the mass per unit area is obtained by the thickness functions $\mathcal{M}_{\mathrm{P}(i,j)} = m_N T_A(r_{\mathrm{P}(i,j)})$ and $\mathcal{M}_{\mathrm{T}(i,j)} = m_N T_A(r_{\mathrm{T}(i,j)})$, while the initial energy density is proportional to the product of the two nuclear thickness functions $\varepsilon_{0 (i,j)} \sim T_A(r_{\mathrm{P}(i,j)})T_A(r_{\mathrm{T}(i,j)})$. We can then solve for the final rapidity $y_{\rm P(i,j)}$ and the nuclear excitation energy $\mathcal{M}_{\mathrm{P}(i,j)}$. The number of baryons within the projectile tube at position $(r_i, \phi_j)$ is $\Delta N_{\mathrm{P}(i,j)} = r_i T_{\mathrm{P}(i,j)} \Delta r\Delta \phi$.  To calculate the baryon rapidity distribution for $b\neq 0$, we perform an integration over the Dirac delta function while fixing the value of $\phi$.
\begin{equation}
\begin{split}
\frac{dN_{\rm P}}{dy} &= \sum_{j}\Delta \phi \int_0^{\infty} d{r_{\perp}} T_{\mathrm{P}}(r_{\perp}, \phi_j, b) r_{\perp} \delta(y-y_{\mathrm{P}}(r_{\perp},\phi_j,b))\\
&=\sum_j\frac{T_{\mathrm{P}}(r_{\perp}, \phi_j, b) r_{\perp}\Delta \phi}{\left|dy_{\mathrm{P}}/dr_{\perp}\right|}\bigg\vert_{r_{\perp} = r_{\perp}(y,\phi_j,b)}\\
\end{split}
\end{equation}

\begin{figure}[thp]
 \centering
 \includegraphics[scale=0.8]{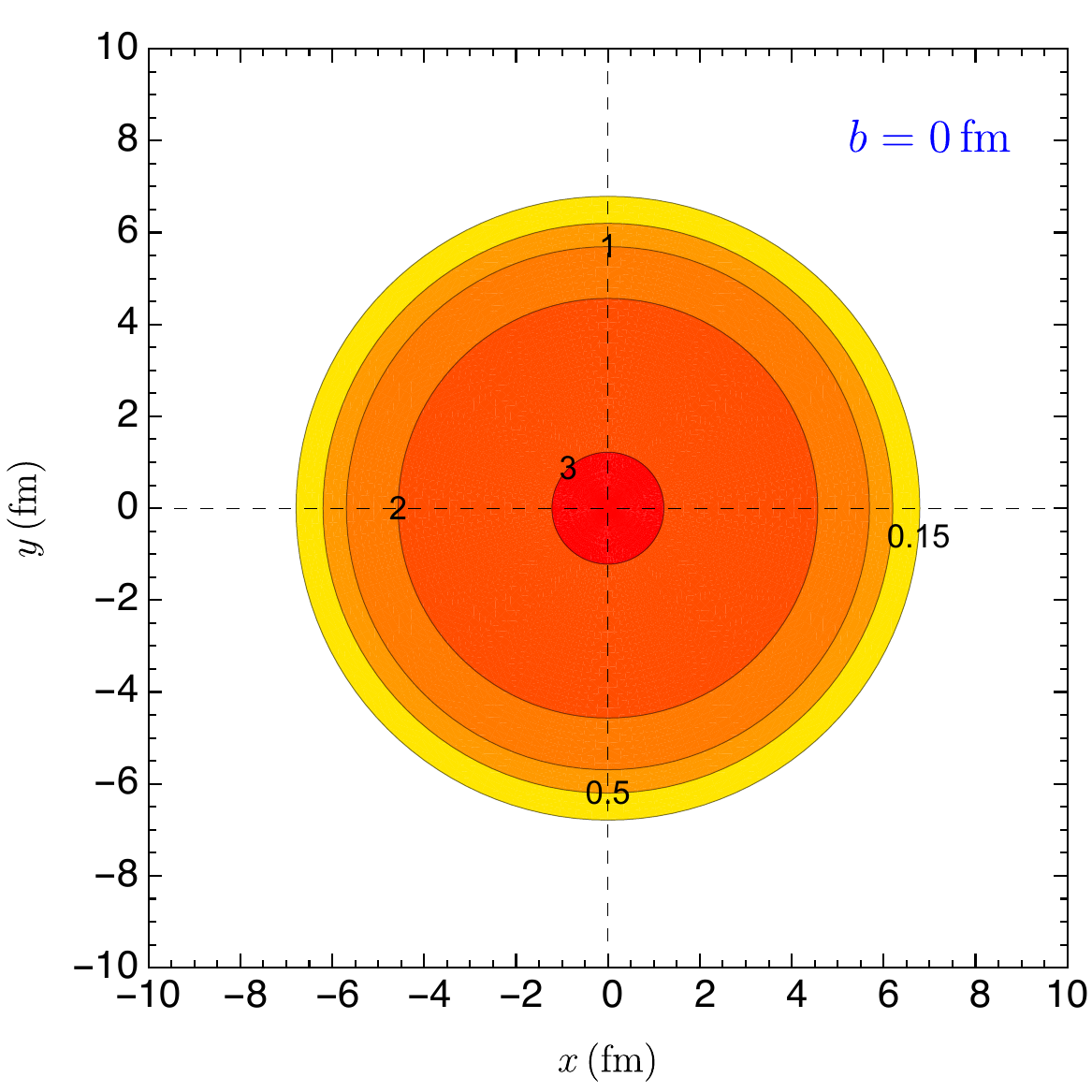}
 \caption{Contour plot of the proper baryon density for the gold \textit{projectile fireball} in the transverse plane for the slice $z^{\prime}=0$.  The 
collision is Au+Au at $\sqrt{s_{NN}} = 200\, \rm{GeV}$ with impact parameter $b=0\,\rm{fm}$. The numbers are in units of baryons per fm$^3$. The position $x=0, y=0$ corresponds to the center of the fireball.  }
 \label{fig:nB_X_Y_b0fm}
\end{figure}

\begin{figure}[thp]
 \centering
 \includegraphics[scale=0.8]{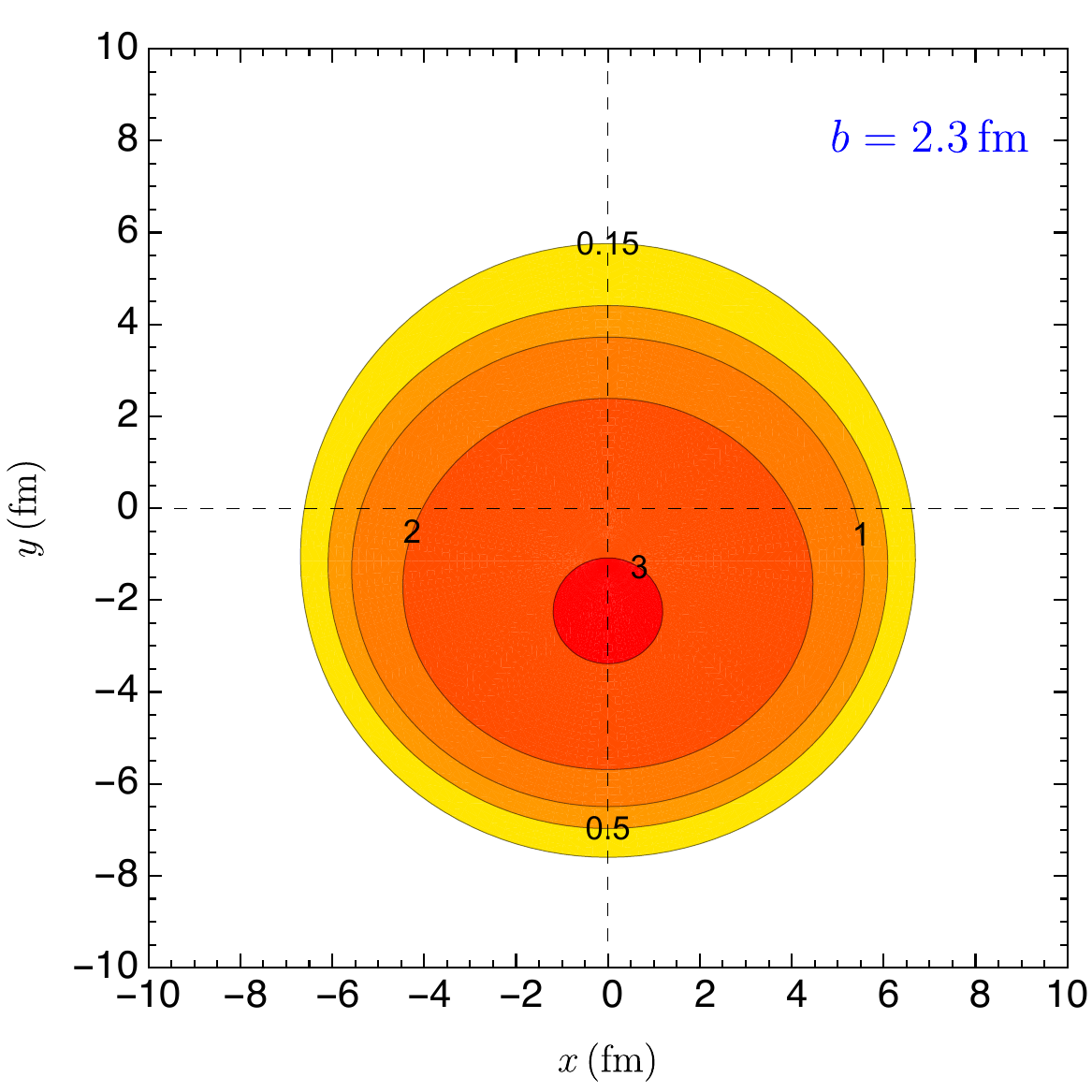}
 \caption{Contour plot of the proper baryon density for the gold \textit{projectile fireball} in the transverse plane for the slice $z^{\prime}=0$. The 
collision is Au+Au at $\sqrt{s_{NN}} = 200\, \rm{GeV}$ with impact parameter $b=2.3\,\rm{fm}$. The numbers are in units of baryons per fm$^3$. The highest density occurs at $x=0, y=-b$.  }
 \label{fig:nB_X_Y_b2p3fm}
\end{figure}

In Figs. \ref{fig:nB_X_Y_b0fm} and \ref{fig:nB_X_Y_b2p3fm} the proper baryon density distribution of the fireball in the transverse plane for the longitudinal slice $z^{\prime}=0$ is given for $b=0$ and $b=2.3\,\rm{fm}$,  respectively.  For zero impact parameter, the baryon distribution is azimuthally symmetric in the transverse plane and the center of the fireball achieves the largest baryon density of about $3\, \rm{baryons/fm}^3$.  In contrast, for collisions with non-zero impact parameter $b=2.3\,\rm{fm}$, the baryon distribution in the transverse plane is no longer azimuthally symmetric. Even the region where the largest baryon density is achieved has been shifted away from the center of the projectile fireball [the center of the fireball corresponds to $(x=0, y=0)$]. The region with the largest baryon density surrounds the point $(x=0, y=2.3\,\rm{fm})$, which is the transverse location of the center of the \textit{target fireball} when the projectile nucleus and the target nucleus overlap. This is easy to understand from our previous discussion on asymmetric Cu+Au collisions. When two nuclear tubes collide, the tube with less baryon charge experiences larger nuclear compression, while the tube with more baryon charge experiences less nuclear compression. 

\begin{figure}[thp]
 \centering
 \includegraphics[scale=0.8]{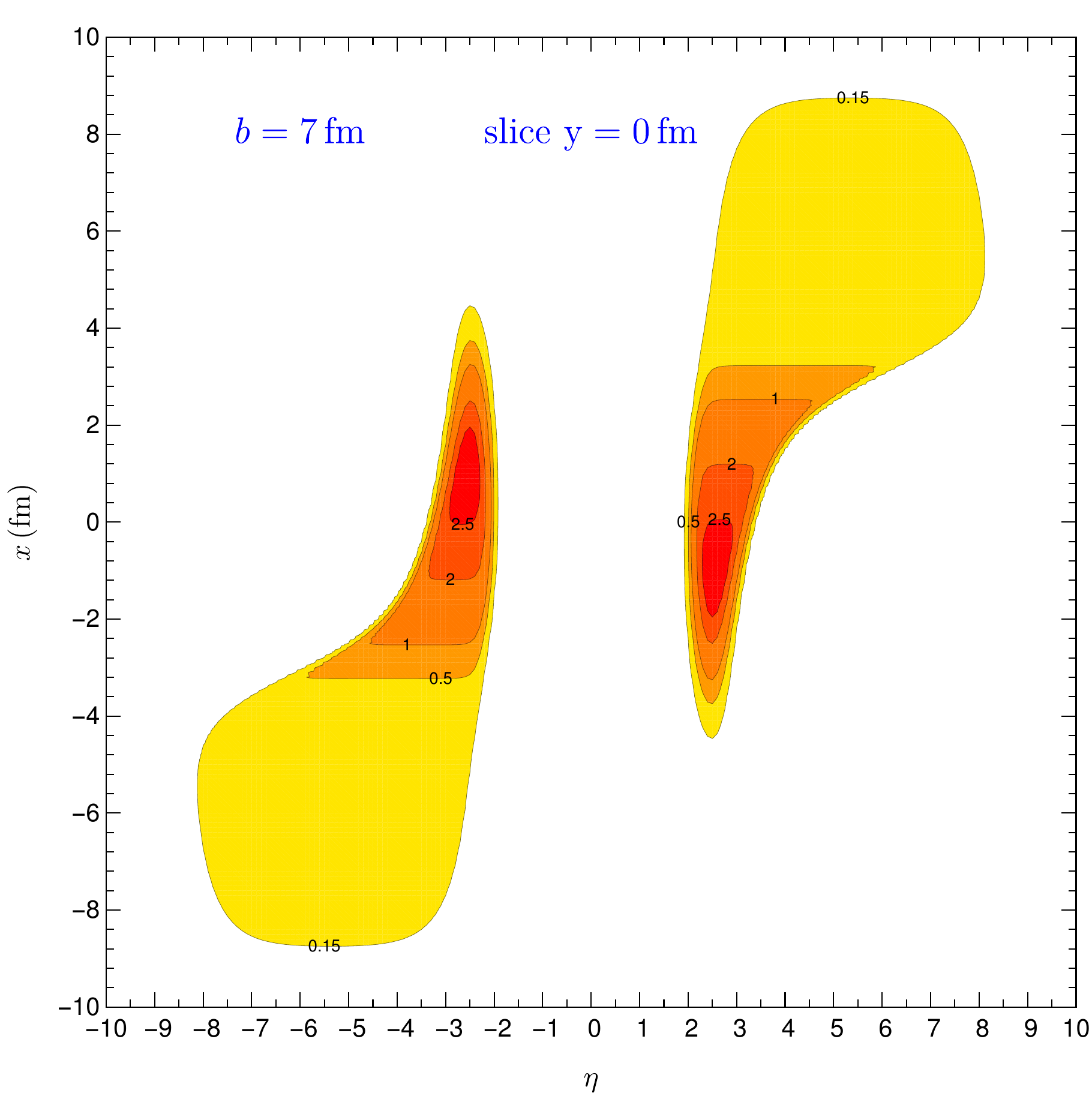}
 \caption{Contour plot of the proper baryon density for Au+Au collisions at $\sqrt{s_{NN}} = 200\, \rm{GeV}$ with impact parameter $b=7\,\rm{fm}$ in the $x$-$\eta$ plane for $y = 0$. The numbers are in units of baryons per fm$^3$. }
\label{baryon_density_etas_x_b=7fm}
\end{figure}

Figure \ref{baryon_density_etas_x_b=7fm} shows the baryon density distribution in the  $x$-$\eta$ plane, where the impact parameter is in the $y$ direction.  The region of the projectile nucleus that overlaps with the central region of the target nucleus experiences the largest nuclear compression while the regions of the projectile nucleus that overlap with the peripheral region of the target nucleus experiences less nuclear compression. 

\section{Conclusion}
\label{section:Conclusion}

In this paper we have systematically studied the high baryon densities outside the central rapidity region of high energy heavy-ion collisions within the McLerran-Venugopalan model. The off-diagonal term in the energy-momentum tensor of the glasma, which comes from the transverse chromo-electromagnetic fields,  is responsible for the nuclear excitation energy in this model. In contrast, typical string models do not deposit energy in the receding nuclei.  For central Au+Au collisions at $\sqrt{s_{NN}} = 200\,\rm{GeV}$, the highest baryon density $3.0\,\rm{baryons/fm}^3$ is about 20 times larger than normal nuclear density, and the largest energy density $20\,\rm{GeV/fm}^3$ is more than 100 times larger than the energy density of nuclear matter at its saturation density. Using a crossover equation of state, the temperature ranges from about 155 to 330 MeV, while and the baryon chemical potential ranges from about 650 to 1020 MeV. The entropy per baryon ratios are found to be in the range of 10 to 23.5, corresponding to the momentum space rapidity range 2.5 to 3.7. The entropy per baryon ratios might be in the right range so that a scan through rapidity may locate the critical point of QCD phase diagram for central collisions.  For central collision at fixed beam energy, the highest baryon density achievable in the fragmentation regions increases with nuclear size. In Cu+Cu, Au+Au and U+U (tip-tip) collisions at $\sqrt{s_{NN}}=200\,\rm{GeV}$, the highest baryon densities are about 2, 3 and 4 baryons/fm$^3$, respectively. In asymmetric Cu+Au collisions, the Cu nucleus is compressed more than the Au nucleus so that the Cu fireball achieves the higher baryon densities.  For central collisions at fixed nuclear size, the highest baryon density achievable increases with collision energy.  We numerically studied Au+Au collisions at $\sqrt{s_{NN}}$ =62.4 and 200 GeV and Pb+Pb collisions at $\sqrt{s_{NN}}$ = 2.76 and 5.02 TeV. For non-central collisions, the average rapidity loss is reduced due to the spectator baryons in the peripheral region of the colliding nuclei. For Au+Au collisions at $\sqrt{s_{NN}}$ = 200 GeV with 0-5\% centrality, our calculations predict the average rapidity loss to be about 2.14. This is in agreement with measurement by the BRAHMS collaboration.  Rapidity loss at LHC is not known because of the challenge of measuring and identifying particles with large rapidity in the detector's frame of rest.  We must emphasize that the results obtained here provide the initial conditions for relativistic fluid dynamic descriptions of high energy heavy ion collisions and cannot be compared directly to experimental data. 

Improvements can be made within the theoretical framework described here.  For example, the McLerran-Venugopalan model for the glasma, as implemented by us, could be complimented by the inclusion of the production of minijets.  An equation of state could be used that incorporates a critical point.  In the presence of a critical point, the adiabatic trajectories as shown in Fig. 14 could be tilted to pass through the critical point if they are within the critical region, see Ref. \cite{Asakawa:2008ti}. In high energy heavy-ion collisions, when scanning the momentum rapidities outside of the central region, the rapidity dependence of the cumulants might be helpful in finding the critical point. Similar rapidity dependence around the central rapidity region in the low energy BES program has recently been proposed in Ref. \cite{Brew}.  For the range of energy and baryon densities where matter in thermodynamic equilibrium ought to be in a mixed phase, it may be that it is initially produced as either a metastable superheated hadronic gas or a metastable supercooled quark-gluon plasma, from which the other phase would have to be nucleated \cite{JK1,JK2,JK3}.  But perhaps the biggest challenge is how to identify and measure baryons (and mesons) at the higher rapidities of relevance to the high baryon density matter.

\newpage

\section*{Acknowledgement}
We are grateful to C. Shen for enlightening discussions.  This work was supported by the U.S. Department of Energy Grant DE-FG02-87ER40328.  ML was also supported by a Doctoral Dissertation Fellowship from the University of Minnesota.


\end{document}